\documentclass[preprint,12pt,authoryear]{elsarticle}

\usepackage{amsmath}
\usepackage{amssymb}
\usepackage{lmodern}
\usepackage{hyperref}
\usepackage{paralist}
\usepackage{natbib}
\usepackage{booktabs}
\usepackage{array}
\usepackage[autostyle]{csquotes}
\usepackage{mathtools}
\usepackage{algorithm}
\usepackage{algorithmic}
\usepackage{bm}
\usepackage{enumitem}
\usepackage{mathrsfs}
\usepackage{float}
\usepackage[font=footnotesize]{caption}
\usepackage{graphicx}
\usepackage{xcolor}

\newtheorem{proposition}{Proposition}
\newtheorem{theorem}{Theorem}
\newcommand{\Sadj}{\hat S^{\mathrm{adj}}_i}
\newcommand{\Sers}{\hat S^{\mathrm{ers}}_i}
\newcommand{\Sun}{\hat S^{\circ}_i}
\newcommand{\ind}{\mathbf{1}}

\journal{Environmental Modelling \& Software}

\ifdefined\pdfgentounicode
  \input{glyphtounicode}
  \def\cmexuni#1#2{\pdfglyphtounicode{#1big}{#2}\pdfglyphtounicode{#1Big}{#2}%
    \pdfglyphtounicode{#1bigg}{#2}\pdfglyphtounicode{#1Bigg}{#2}}
  \cmexuni{parenleft}{0028}\cmexuni{parenright}{0029}
  \cmexuni{bracketleft}{005B}\cmexuni{bracketright}{005D}
  \cmexuni{braceleft}{007B}\cmexuni{braceright}{007D}
  \cmexuni{floorleft}{230A}\cmexuni{floorright}{230B}
  \cmexuni{ceilingleft}{2308}\cmexuni{ceilingright}{2309}
  \cmexuni{angbracketleft}{27E8}\cmexuni{angbracketright}{27E9}
  \cmexuni{slash}{002F}\cmexuni{backslash}{005C}\cmexuni{radical}{221A}
  \pdfglyphtounicode{integraltext}{222B}\pdfglyphtounicode{integraldisplay}{222B}
  \pdfglyphtounicode{summationtext}{2211}\pdfglyphtounicode{summationdisplay}{2211}
  \pdfglyphtounicode{producttext}{220F}\pdfglyphtounicode{productdisplay}{220F}
  \pdfglyphtounicode{uniontext}{22C3}\pdfglyphtounicode{uniondisplay}{22C3}
  \pdfglyphtounicode{intersectiontext}{22C2}\pdfglyphtounicode{intersectiondisplay}{22C2}
  \pdfglyphtounicode{vextendsingle}{2223}\pdfglyphtounicode{vextenddouble}{2225}
  \pdfglyphtounicode{mapsto}{007C}
\fi

\begin{document}

\begin{frontmatter}

\title{An Adjusted Ersatz Discrepancy for Global Sensitivity Analysis: Theory and Benchmarks}

\author[label1,label2]{Samuele Lo Piano}
\affiliation[label1]{organization={Faculty of Management and Economics, Gdańsk University of Technology},
             city={Gdańsk},
             country={Poland}}
\affiliation[label2]{organization={Barcelona School of Management, Pompeu Fabra University},
             city={Barcelona},
             country={Spain}}

\author[label3]{Alessio Lachi\corref{cor1}}
\ead{a.lachi@unilink.it}
\cortext[cor1]{Corresponding author.}
\affiliation[label3]{organization={Department of Life Sciences, Health, and Health Professions, Link Campus University},
             city={Rome},
             country={Italy}}

\author[label4]{Razi Sheikholeslami}
\affiliation[label4]{organization={Department of Earth and Atmospheric Sciences, University of Alberta},
             city={Edmonton},
             state={AB},
             country={Canada}}

\author[label5]{Arnald Puy}
\affiliation[label5]{organization={School of Geography, Earth and Environmental Sciences, University of Birmingham},
             city={Birmingham},
             country={United Kingdom}}

\author[label6]{Pamphile Tupui Roy}
\affiliation[label6]{organization={Consulting Manao},
             city={Vienna},
             country={Austria}}

\author[label2,label7]{Andrea Saltelli}
\affiliation[label7]{organization={Centre for the Study of the Sciences and the Humanities, University of Bergen},
             city={Bergen},
             country={Norway}}

\begin{abstract}
In given-data sensitivity analysis the analyst holds only a fixed sample of model inputs and outputs. We revisit the ersatz discrepancy of Puy, Roy and Saltelli (2024), a grid-occupancy statistic read from the scatterplot of an input against the output, and propose two modifications: a rank transformation of the output and a deterministic Moore-neighbourhood fill of isolated empty cells. We derive the asymptotic theory of the resulting adjusted ersatz under simple random sampling. Without the fill, the rank-transformed statistic is exactly a histogram estimator of Borgonovo's moment-independent measure $\delta$. At its resolution, however, it converges to $e^{-1}$ plus a different $f$-divergence, so that its value under independence is $0.37$. With the fill, the limit lies between $\delta-0.141$ and $\delta+0.048$ for every dependence structure, and its value under independence drops to $0.047$. The theory explains the type~I errors observed at screening thresholds of $0.01$, $0.05$ and $0.10$, the dominant role of the grid resolution, and when rankings by $\delta$-type measures and by total-order Sobol' indices agree. In replicated benchmarks, Becker metafunctions and the HYMOD model, a degree-adaptive polynomial chaos expansion estimates total-order indices most accurately, whereas the adjusted ersatz differs from $\delta$ by $0.02$--$0.08$ on average, as the theory predicts, and improves markedly on the original measure. It is best used as a fast, assumption-free screening statistic.
\end{abstract}


\end{frontmatter}


\section{Introduction}

Global sensitivity analysis (GSA) investigates how the uncertainty in model inputs, varied simultaneously over their full ranges, propagates to the model outputs. It helps modellers test assumptions, identify the key drivers of a model, simplify it and support decisions based on its results \citep{Saltelli2008}.

Over the past three decades many GSA techniques have been developed, with extensive guidance on their selection \citep{becker_metafunctions_2020, puy_comprehensive_2022, razavi2021future, sheikholeslami2021viscous}. Variance-based methods decompose the output variance of a model with $k$ inputs into contributions of individual inputs, pairs, triplets and higher-order interactions. The first-order index $S_i$ measures the direct effect of input $i$, and the total-order index $T_i$ its direct effect together with all its interactions \citep{homma1996importance}. $T_i$ has a direct operational meaning for screening: fixing input $i$ at a random value changes the output by a mean squared error equal to $2\,T_i\,\mathrm{Var}(Y)$ \citep{sobol2001}, so that $T_i=0$ characterises the inputs that can be fixed without loss.

$T_i$ is usually estimated by Monte Carlo integration on structured designs, for instance with the Jansen estimator \citep{jansen1999, saltelli_variance_2010}, or from an emulator such as a polynomial chaos expansion \citep[PCE;][]{sudret_global_2008} or a machine-learning model \citep{antoniadis2021random}. Shapley effects allocate the output variance among the inputs following the Shapley value of cooperative game theory. For independent inputs they sum to one and satisfy $S_i\le\mathrm{Sh}_i\le T_i$ \citep{owen2014}.

In the given-data setting the analyst holds a fixed sample of input--output pairs and cannot design further model runs \citep{plischke2013, daveiga2015, pianosiwagener2018}. Emulator-based estimators then rest on a functional-form assumption, whose adequacy has to be checked on the same sample. Distribution-based measures avoid this assumption. They quantify how much the conditional distribution of the output given an input departs from its unconditional distribution, rather than how much output variance the input explains. Examples are the $\delta$ measure of \citet{borgonovo2007}, the $f$-sensitivity indices of \citet{rahman2016} and the PAWN index of \citet{pianosiwagener2015}. Variance is not always an adequate summary of output uncertainty, for instance for skewed or multimodal outputs \citep{borgonovo2011}. Variance-based and distribution-based measures can also be combined on the same sample. In the strategy of \citet{baroni2020}, the variance of the conditional means of the output measures the first-order effect of an input, and Borgonovo's $\delta$ computed between conditional distributions of the output centred at their means measures its interactions. With this index they found interactions that the variance-based difference $T_i-S_i$ missed, in particular when the output distribution was skewed.

\citet{puy2024discrepancy} showed that discrepancy measures computed on the plane $[x_i,Y]$, where $x_i$ is an input and $Y$ the model output rescaled to the unit interval, reproduce the ranking of $T_i$ well, and proposed a cheap substitute for them, the S-ersatz. The S-ersatz divides the scatterplot of $x_i$ against the scaled output into a grid of about as many cells as sample points and counts the cells that contain points: an input without influence leaves the points spread over the plane, whereas an influential input concentrates them along a curve and leaves many cells empty (Figure~\ref{fig:scatter}: $x_5$ to $x_8$ against $x_1$).

Two elements, however, are missing for the S-ersatz to serve as a trustworthy screening tool. First, no theory states what it estimates and when its ranking can be trusted. Second, it has not been compared, at equal sample cost, with the estimators it could replace in the given-data setting.

This paper supplies both, together with an adjustment of the measure. It makes four contributions.
\begin{enumerate}[label=\roman*),leftmargin=*,itemsep=0pt,topsep=2pt]
    \item We introduce the adjusted ersatz, which rank-transforms the output before gridding and fills isolated empty cells with a Moore-neighbourhood rule (\S\ref{1st} and \S\ref{2nd}).
    \item We derive an asymptotic theory of the ersatz statistics under simple random sampling (SRS), which states what they estimate (\S\ref{theory} and Supplementary Material A).
    \item We compare the adjusted ersatz, at equal sample cost, with three given-data estimators (\S\ref{comparators}, \S\ref{design} and Section~\ref{results}).
    \item We identify the algorithmic choices that matter through a joint sensitivity analysis of the method's own parameters (\S\ref{SoS}).
\end{enumerate}
The theory shows that the rank-transformed grid without fill is exactly a histogram estimator of $\delta$, which at the resolution of the ersatz converges to $e^{-1}$ plus a different $f$-divergence, and that with the fill the limit stays within a fixed distance of $\delta$ for every dependence structure. It also gives the value of the statistic under independence, which explains its type~I errors, the role of the grid resolution, and conditions under which rankings by $\delta$-type measures and by $T_i$ agree. The comparators represent the approaches reviewed above: total-order indices and Shapley effects from a PCE, and the given-data PAWN index. They are assessed against two references, $T_i$ and $\delta_i$, with replicated samples and three sampling designs, on three aspects of performance: the agreement of the rankings, measured by the correlation of Savage scores (which gives most weight to the most influential inputs), the distance of the values from the references, measured by the mean absolute error, and the screening errors, measured by the type~I and type~II error rates at three thresholds.

Section~\ref{methods} introduces the statistics, their theory, the comparator estimators, and the experimental design with the performance metrics. Section~\ref{results} reports the convergence study, the benchmark functions, the Becker metafunction ensemble, the sensitivity analysis of the method and the HYMOD application. Section~\ref{disc} discusses the results and concludes.

\section{Methods}\label{methods}

\subsection{Notation and what a scatterplot shows}\label{scatter}

Let $\bm X=(X_1,\dots,X_k)$ be the vector of independent model inputs and $Y=f(\bm X)$ the scalar model output. A given-data sample consists of an $N\times k$ matrix $\mathbf X$ with rows $\bm x^{(j)}=(x_1^{(j)},\dots,x_k^{(j)})$ and the vector $\bm y=(y_1,\dots,y_N)$ of outputs $y_j=f(\bm x^{(j)})$, $j=1,\dots,N$. The inputs are expressed on the probability scale, $x_i^{(j)}\in[0,1]$, by applying the cumulative distribution function of each input. All estimators below operate on this scale.

\begin{figure}[H]
    \centering
    \includegraphics[width=\textwidth]{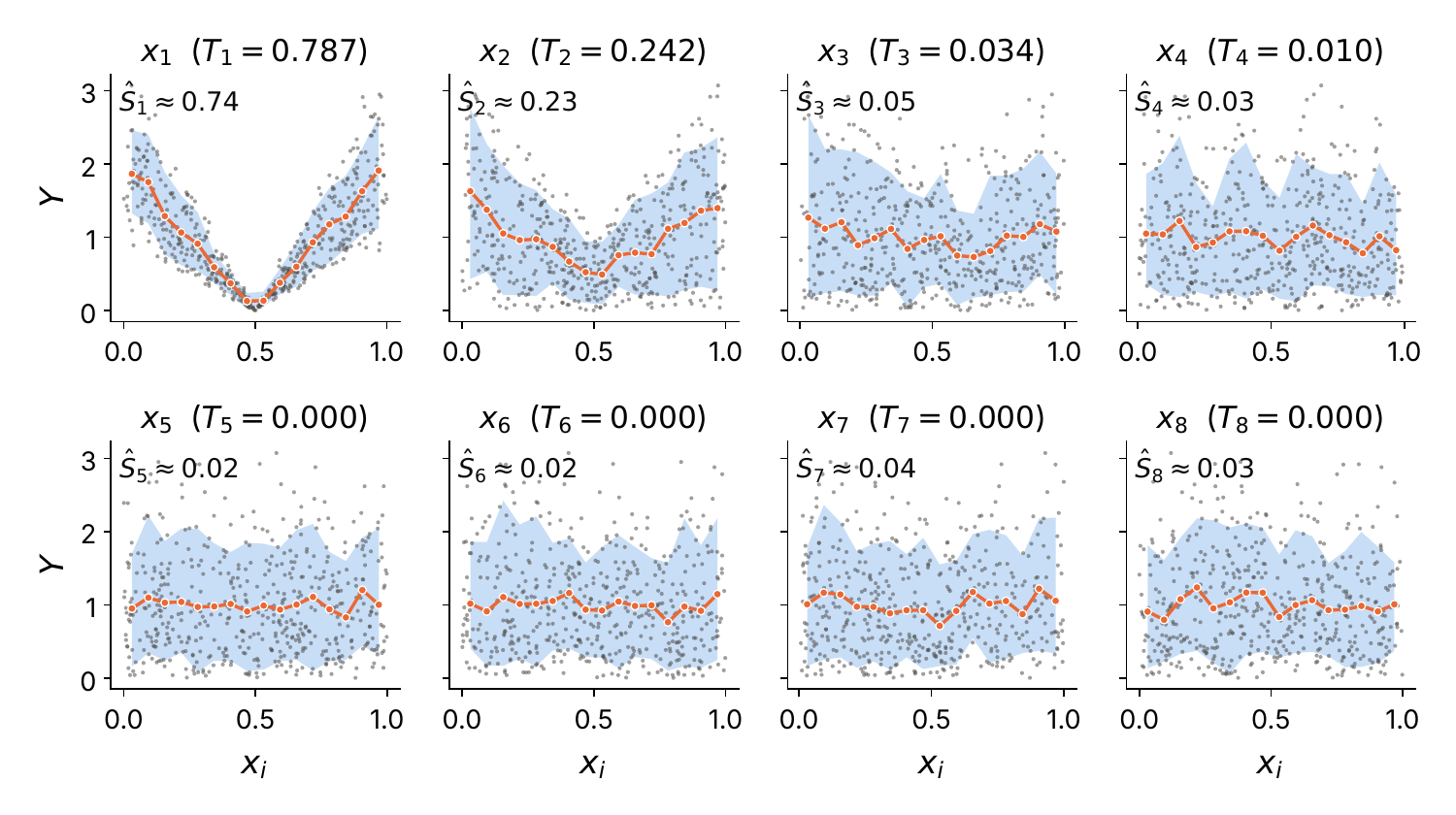}
    \caption{Scatterplots of the Sobol' $G$ function (Table~\ref{tab:datasets}) for a simple random sample of size $N=2^9$. Orange dots: conditional means of $Y$ in 16 equal-width bins of $x_i$, with $\hat S_i$ the ratio of their variance to the output variance, a rough first-order estimate. Shaded band: 10\% to 90\% conditional quantiles. The conditional mean carries the first-order effect, and the change of the conditional spread with $x_1$ and $x_2$ is the imprint of their interactions with the other inputs.}
    \label{fig:scatter}
\end{figure}

Scatterplots of $\bm y$ against each column of $\mathbf X$ are a common first step of a sensitivity analysis. Figure~\ref{fig:scatter} shows them for the Sobol' $G$ function, a product of one factor per input, and two of their features are informative. The conditional mean (orange dots) carries the first-order effect: $S_i>0$ if and only if it varies with $x_i$, as for $x_1$ and $x_2$. The conditional spread, the width of the shaded band, does not change with $x_i$ when the output is additive in $x_i$ (Proposition~\ref{prop:sobol} below), so a band of varying width reveals interactions of $x_i$ with the other inputs. For $x_1$ and $x_2$ the band narrows at $x_i=1/2$, where their factors, which scale the contribution of all the other inputs, are smallest. For $x_5$ to $x_8$ ($T_i\approx10^{-4}$) neither feature varies beyond sampling noise. The spread can reveal an interaction without a first-order effect: for $Y=X_1X_2X_3$ with inputs centred at zero, $S_1=0$, but the band widens with $|x_1|$.

\citet{baroni2020} quantify the two features separately: the first through the variance of the conditional means, the second through Borgonovo's $\delta$ between conditional distributions of the output centred at their means. Proposition~\ref{prop:sobol}(c) gives a formal basis to the second index. When the output is additive in $x_i$, the centred conditional distributions coincide, so the index vanishes, and a positive value is evidence of interactions. The converse does not hold. Interactions that leave the conditional distribution of the output unchanged, as in $Y=(X_1+X_2)\bmod1$, are not visible in a scatterplot at all (Proposition~\ref{prop:sobol}(b)). Any statistic computed from the pairs $(x_i^{(j)},y_j)$, including the ones studied here, shares this limitation. It measures the dependence between $X_i$ and $Y$, which comprises the direct influence of $X_i$ and the imprint of its interactions of any order on the conditional distribution of the output, but does not exhaust the contributions to $T_i$.

\subsection{The S-ersatz}\label{ersatz}

\citet{puy2024discrepancy} partition the plane of $x_i$ against the min-max scaled output into an $s\times s$ grid of square cells, with $s=\lceil\sqrt N\rceil$ and $\lceil\cdot\rceil$ the ceiling function (for $N=16$, $s=4$; for $N=32$, $s=6$), and define the S-ersatz as the fraction of occupied cells. We use its complement, so that, as for $T_i$, larger values indicate a more influential input. With $n_{ab}$ the number of points in cell $(a,b)$, the S-ersatz is then the fraction of empty cells,
\begin{equation}\label{eq:sersatz}
\Sers=1-\frac{1}{s^2}\sum_{a=1}^{s}\sum_{b=1}^{s}\ind[n_{ab}>0],
\end{equation}
where $\ind[\cdot]$ is the indicator function. In the $4\times4$ example
\begin{equation*}
    \begin{bmatrix}
        \bullet  & \circ   & \bullet & \bullet \\
        \bullet  & \bullet & \bullet & \circ   \\
        \circ    & \bullet & \circ   & \circ   \\
        \circ    & \circ   & \circ   & \circ
    \end{bmatrix},
\end{equation*}
where $\bullet$ and $\circ$ denote occupied and empty cells, seven of the sixteen cells are occupied and $\Sers=1-7/16=0.5625$. An input without influence leaves the points spread over the plane and the statistic low, whereas a strongly influential input concentrates them and empties many cells. The name refers to the use of the statistic as a substitute for discrepancy measures, which quantify how far a set of points is from the uniform distribution on the unit square. Because the output is min-max scaled rather than rank-transformed, neither the S-ersatz nor the discrepancy measures of \citet{puy2024discrepancy} depend on $(X_i,Y)$ through its copula alone. They compare the joint distribution of $X_i$ and the scaled output with the uniform distribution on the square, which is the independence copula, but that joint distribution is the copula of $(X_i,Y)$ only when the scaled output is uniform. Like the correlation ratio, they are invariant under increasing affine transformations of the output but not under other monotonic transformations. They are therefore not copula-based in the sense of \citet[Table~1]{plischke2019}, which requires invariance under monotonic transformations and an expression in terms of the copula. The rank transformation of Section~\ref{1st} removes this dependence on the distribution of the output.

\subsection{First modification: rank transformation of the output}\label{1st}

\begin{figure}[H]
    \centering
    \includegraphics[width=\textwidth]{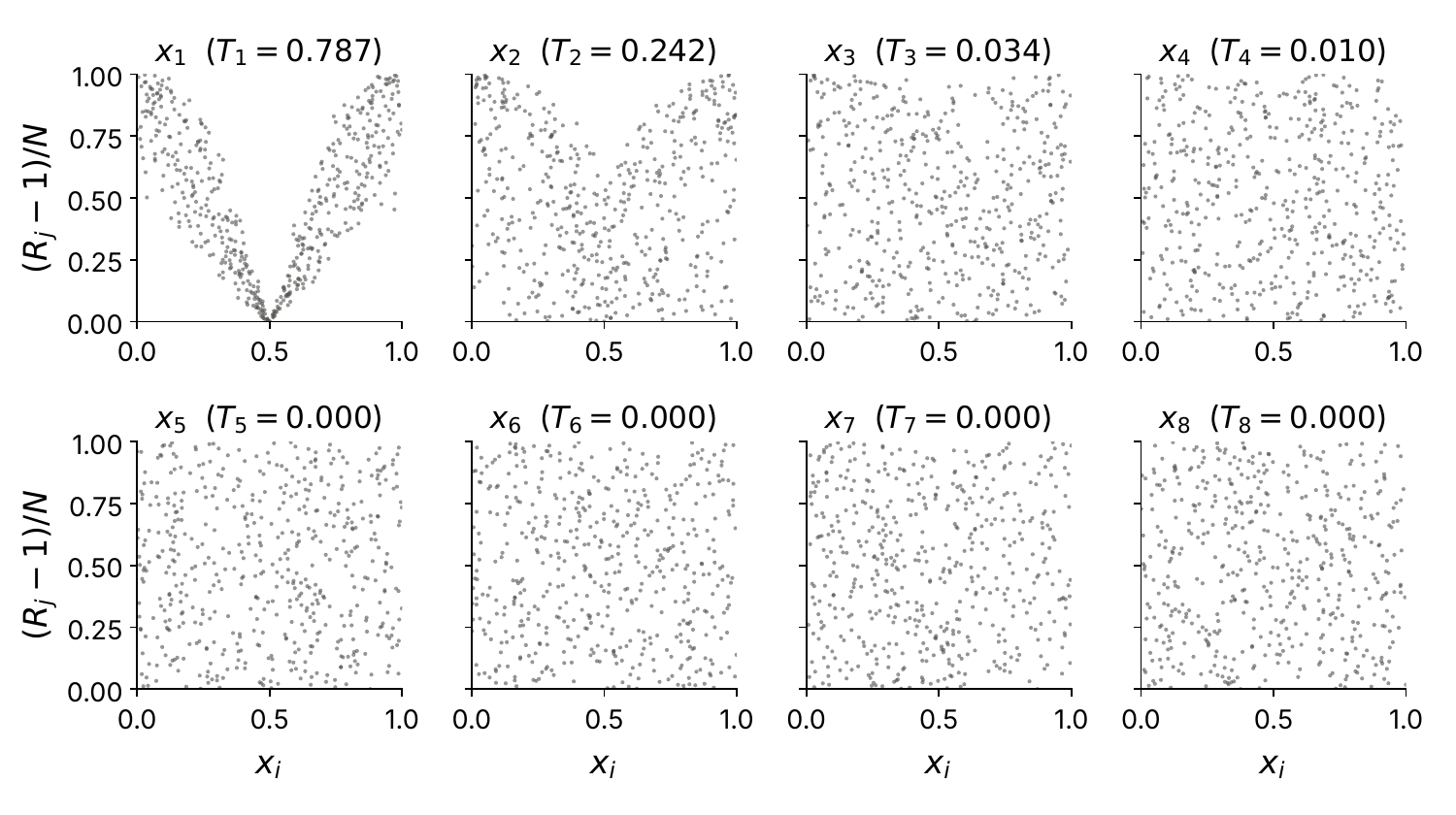}
    \caption{Rank plots (scatterplots of the pseudo-observations $(x_i^{(j)},v_j)$, Equation~\eqref{eq:ranks}) for the sample of Figure~\ref{fig:scatter}.}
    \label{fig:rankplot}
\end{figure}

\begin{figure}[H]
    \centering
    \includegraphics[width=\textwidth]{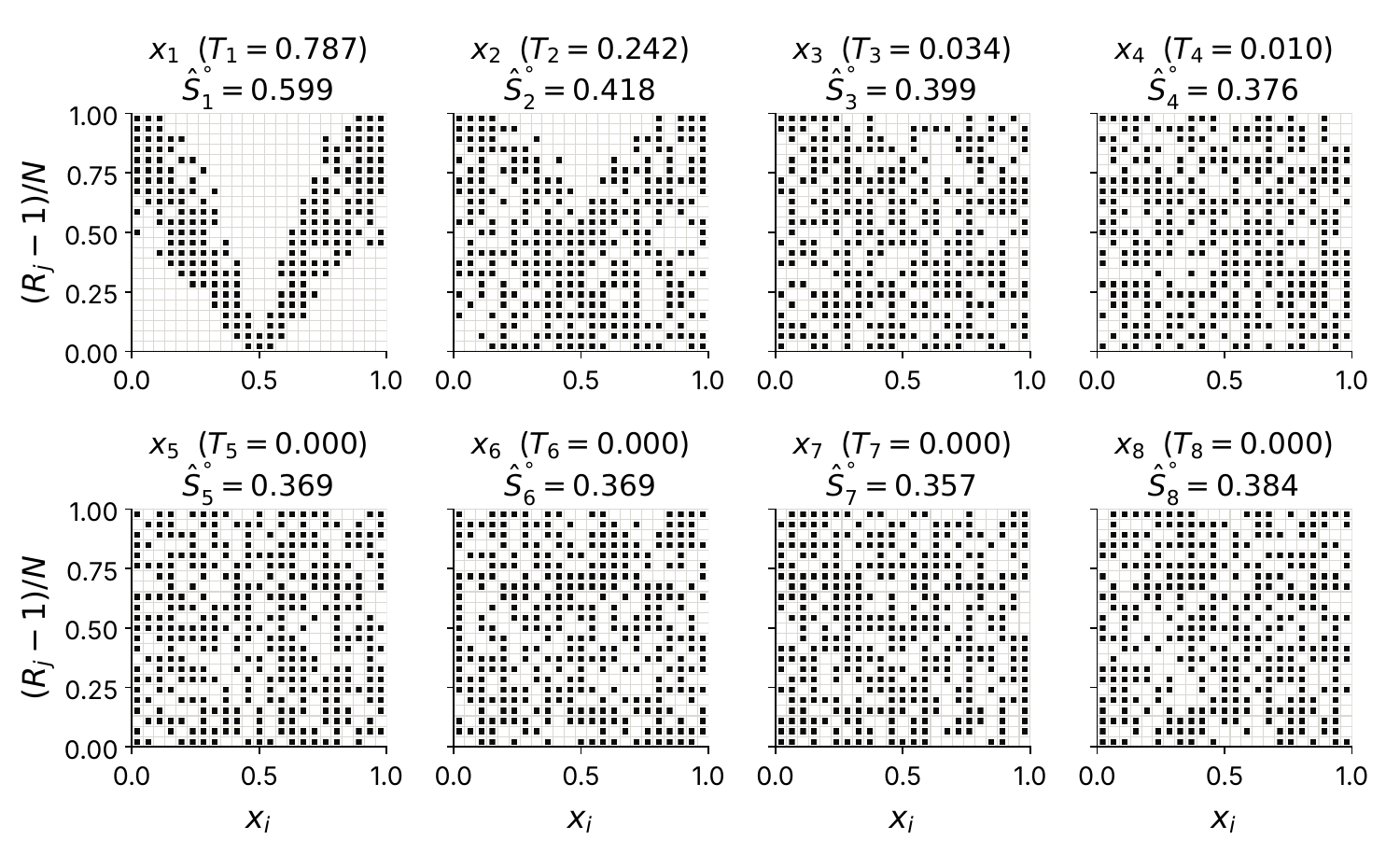}
    \caption{Occupancy grids ($s=23$) of the rank plots of Figure~\ref{fig:rankplot}. Black squares: occupied cells. The share of empty cells, given above each panel, is the rank-transformed statistic without fill, $\Sun$.}
    \label{fig:grid}
\end{figure}

Even when the output does not depend on $x_i$, the S-ersatz depends on the distribution of the output: where the output density is low, entire columns of the grid stay empty. Under independence and SRS the S-ersatz converges to $\int_0^1\exp\{-\tilde f(v)\}\,\mathrm dv$, where $\tilde f$ is the density of the min-max scaled output. This limit exceeds $e^{-1}$, its value for a uniform output, unless the output is uniform, and it tends to one for outputs with unbounded support (Supplementary Material A, Proposition~5.6). We therefore replace the output by its normalised ranks
\begin{equation}\label{eq:ranks}
v_j=\frac{R_j-1}{N},
\end{equation}
where $R_j$ is the rank of $y_j$ among $y_1,\dots,y_N$. Ties are broken by order of appearance in the sample. In a simple random sample this is equivalent to breaking them at random (Supplementary Material A, Lemma~3.1). The pairs $(x_i^{(j)},v_j)$ are pseudo-observations, that is a sample from the empirical copula of $(X_i,Y)$, and their scatterplot is the rank plot of Figure~\ref{fig:rankplot} \citep{genest2007}. Every column of the grid now contains $N/s$ points up to rounding, and the statistic depends on the joint distribution of $(X_i,Y)$ only through its copula. The step is the one that turns the correlation ratio into the rank correlation ratio, which \citet[Section~3.1]{plischke2019} identify with the rank-transformed sensitivity indices of \citet{saltelli1995rank}. In the classification of \citet{plischke2019} the statistic then belongs to niveau~3, which compares the copula density with that of the independence copula and contains $\delta$ (their Equation~30), rather than to niveau~1, where discrepancies in the strict sense compare the copula itself with the independence copula \citep{morokoff1994}. The same transformation would turn the discrepancy measures of \citet{puy2024discrepancy} into niveau~1 measures: the star $L_2$ discrepancy of the pseudo-observations, multiplied by $\sqrt{90}$, is the estimator of Hoeffding's omega statistic, whose square \citet[Equation~39]{plischke2019} compute with Warnock's formula, and it underlies the Cram\'er--von Mises tests of independence of \citet{genest2004}. Figure~\ref{fig:grid} shows the resulting occupancy grids.

\subsection{Second modification: Moore-neighbourhood fill}\label{2nd}

\begin{figure}[H]
    \centering
    \includegraphics[width=\textwidth]{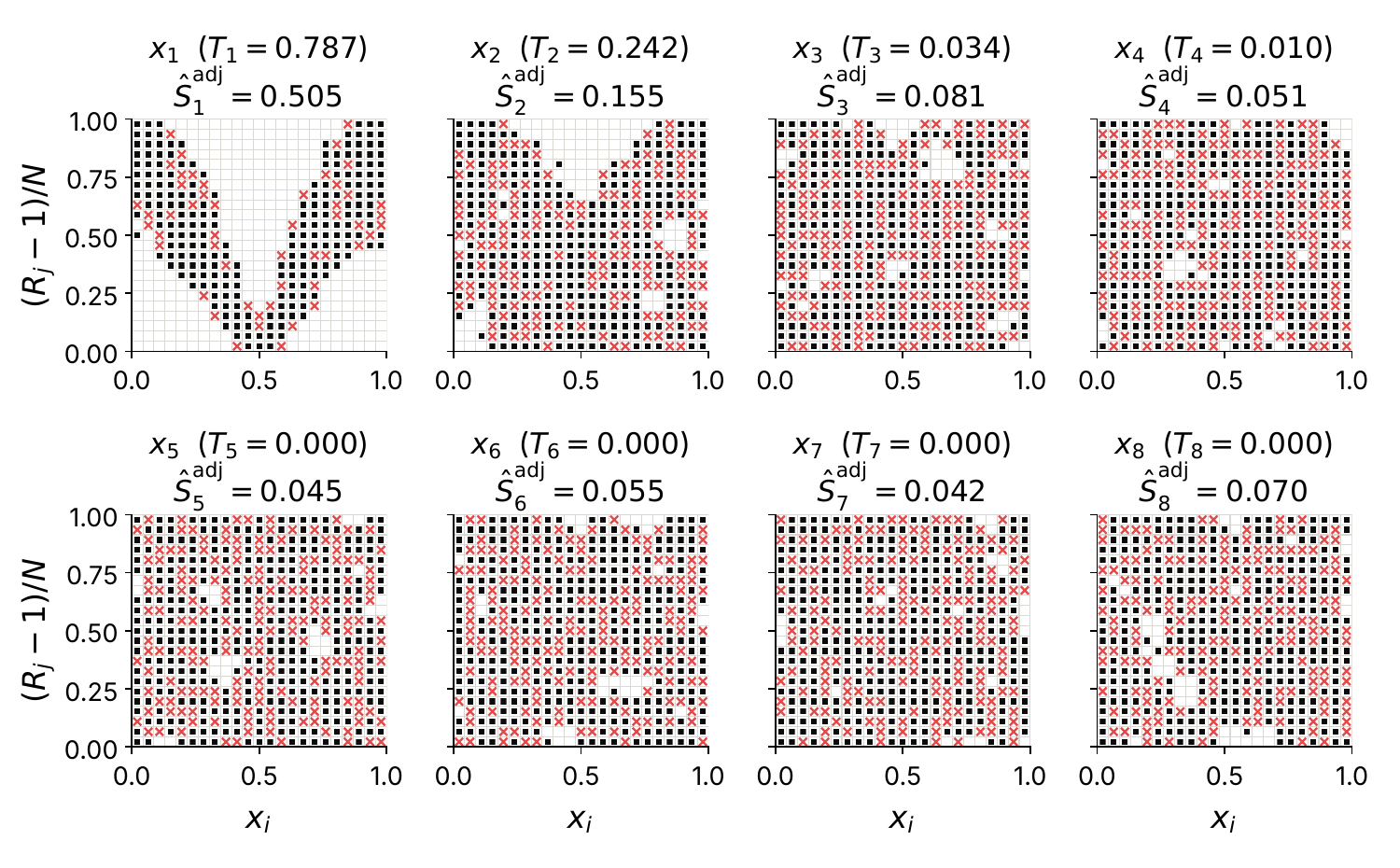}
    \caption{Moore-neighbourhood fill of the grids of Figure~\ref{fig:grid}. Black squares: occupied cells; red crosses: filled cells. The share of cells that remain empty, given above each panel, is the adjusted ersatz $\Sadj$.}
    \label{fig:fill}
\end{figure}

An empty cell in Figure~\ref{fig:grid} may reflect a genuine absence of probability mass or sampling variability: with about one point per cell, about $e^{-1}\approx37\%$ of the cells are empty even when $x_i$ and $Y$ are independent. We therefore fill isolated empty cells. The Moore neighbourhood of a cell is the set of its (at most eight) adjacent cells, diagonal neighbours included, a notion used in cellular automata such as Conway's Game of Life \citep{gardner_mathematical_1970}. An empty cell is filled if at least a fraction $\theta_f$ of its Moore neighbours is occupied, with $\theta_f=1/2$ by default: at least four of eight neighbours for an interior cell, three of five on an edge and two of three in a corner. Neighbours are read on the original grid, so the fill does not cascade. Filling can only reduce the number of empty cells, so the adjusted statistic is never larger than the unfilled one (Supplementary Material A, Proposition~3.2). Its effect is to remove the empty cells produced by sampling variability in regions of positive probability density while keeping the empty cells inside empty regions (Figure~\ref{fig:fill}). The adjusted ersatz is the fraction of cells that remain empty,
\begin{equation}\label{eq:sadj}
\Sadj=1-\frac{1}{s^2}\sum_{a=1}^{s}\sum_{b=1}^{s}\tilde G_{ab},
\end{equation}
where $\tilde G$ is the occupancy matrix after the fill. Algorithm~\ref{alg:s_ersatz_adj} gives the complete computation.

\begin{algorithm}[H]
\caption{Adjusted ersatz $\Sadj$ of input $i$.}
\label{alg:s_ersatz_adj}
\begin{algorithmic}[1]
\REQUIRE inputs $x_i^{(j)}\in[0,1]$ and outputs $y_j$, $j=1,\dots,N$; grid exponent $\alpha_g$ (default $1/2$); fill threshold $\theta_f$ (default $1/2$)
\STATE $s\leftarrow\lceil N^{\alpha_g}\rceil$
\STATE row of each point: $a_j\leftarrow\max\{1,\lceil s\,x_i^{(j)}\rceil\}$
\STATE rank $R_j$ of $y_j$ (ties by order of appearance) and column $b_j\leftarrow\max\{1,\lceil s\,(R_j-1)/N\rceil\}$
\STATE occupancy: $G_{ab}\leftarrow1$ if $(a_j,b_j)=(a,b)$ for some $j$, else $G_{ab}\leftarrow0$
\FOR{every cell $(a,b)$}
\STATE $\tilde G_{ab}\leftarrow1$ if $G_{ab}=1$ or if the mean of $G$ over the in-grid Moore neighbours of $(a,b)$ is at least $\theta_f$; otherwise $\tilde G_{ab}\leftarrow0$
\ENDFOR
\RETURN $\Sadj\leftarrow1-s^{-2}\sum_{a,b}\tilde G_{ab}$
\end{algorithmic}
\end{algorithm}

The cost is $O(N\log N+s^2)$ per input, dominated by the ranking of the output, which is shared by all inputs.

\subsection{What the adjusted ersatz estimates}\label{theory}

This section summarises the asymptotic theory of the statistics. Statements in full, proofs and numerical verifications are given in Supplementary Material A, which marks the steps that rest on a numerical verification of an explicit inequality, such as the upper bound of Proposition~\ref{prop:limits}(c), as computer-assisted. The sample is drawn by SRS: the rows of $\mathbf X$ are independent and uniformly distributed on $[0,1]^k$. $C_i$ denotes the copula of $(X_i,Y)$, $c_i$ the density of its absolutely continuous part and $\Pi$ the independence copula \citep{nelsen2006}. The regularity required of $c_i$ allows unbounded densities, densities with jumps along curves and singular parts concentrated on curves. In copula form \citep{wei2014copula, plischke2019}, the moment-independent measure of \citet{borgonovo2007} is the total-variation distance between $C_i$ and $\Pi$,
\begin{equation}\label{eq:delta}
\delta_i=\frac12\,\mathbb E_{X_i}\!\int\big|f_Y(y)-f_{Y|X_i}(y)\big|\,\mathrm dy=\iint\big(1-c_i(u,v)\big)_+\,\mathrm du\,\mathrm dv .
\end{equation}

\begin{proposition}[histogram identity]\label{prop:hist}
Let $\Sun$ be the statistic without fill and $n_{ab}$ its cell counts. If $N\le s^2$, as for $s=\lceil\sqrt N\rceil$, then for every sample and every design
\[
\Sun=\frac12\sum_{a,b}\Big|\frac{n_{ab}}{N}-\frac1{s^2}\Big|,
\]
the plug-in histogram estimator of $\delta_i$ on the $s\times s$ grid.
\end{proposition}

\begin{theorem}[limits]\label{thm:limits}
Let $s\to\infty$ with $N/s^2\to\lambda>0$ ($\lambda=1$ for $s=\lceil\sqrt N\rceil$), and let $c_i$ satisfy the regularity condition of Supplementary Material A (Assumption~2.1). Under SRS, almost surely (in probability if the output has atoms and ties are broken by order of appearance),
\begin{gather*}
\Sun\to\Phi_\lambda(C_i)=\iint e^{-\lambda c_i(u,v)}\,\mathrm du\,\mathrm dv,\\
\Sadj\to\Psi_\lambda(C_i)=\iint h_\lambda\big(c_i(u,v)\big)\,\mathrm du\,\mathrm dv,
\end{gather*}
where $h_\lambda(t)=e^{-\lambda t}\,\Pr[\mathrm{Bin}(8,1-e^{-\lambda t})<8\theta_f]$ is the probability that a cell stays empty after the fill when the counts of the cell and of its eight neighbours are independent Poisson variables with mean $\lambda t$. The probability of a deviation of the statistics from their conditional means decays exponentially in $N$.
\end{theorem}

\begin{proposition}[the limit functionals]\label{prop:limits}\leavevmode
\begin{enumerate}[label=(\alph*),leftmargin=*,itemsep=1pt]
\item $\Phi_\lambda(C_i)-e^{-\lambda}$ is the $f$-divergence of $C_i$ from $\Pi$ with $f(t)=e^{-\lambda t}-e^{-\lambda}$, that is, an $f$-sensitivity index in the sense of \citet{rahman2016}. It equals zero if and only if $X_i$ and $Y$ are independent, and $\Phi_\lambda(C_i)=1$ if and only if $C_i$ is singular. For every copula, $\Phi_1(C_i)-e^{-1}\le(1-e^{-1})\,\delta_i$, and the constant is sharp. If $C_i$ is absolutely continuous with $c_i\le\bar c$, then $\Phi_1(C_i)-e^{-1}\ge2e^{-\bar c}\,\delta_i^2$.
\item For $\theta_f=1/2$ and $\lambda\ge3/4$, $\Psi_\lambda(C_i)\ge\kappa_\lambda=h_\lambda(1)$, with equality if and only if $X_i$ and $Y$ are independent ($\kappa_1=0.047$), and $\Psi_\lambda(C_i)=1$ if and only if $C_i$ is singular.
\item For $\lambda=1$ and every copula, $\delta_i-0.1401\le\Psi_1(C_i)\le\delta_i+0.0471$.
\end{enumerate}
\end{proposition}

\begin{figure}[H]
    \centering
    \includegraphics[width=\textwidth]{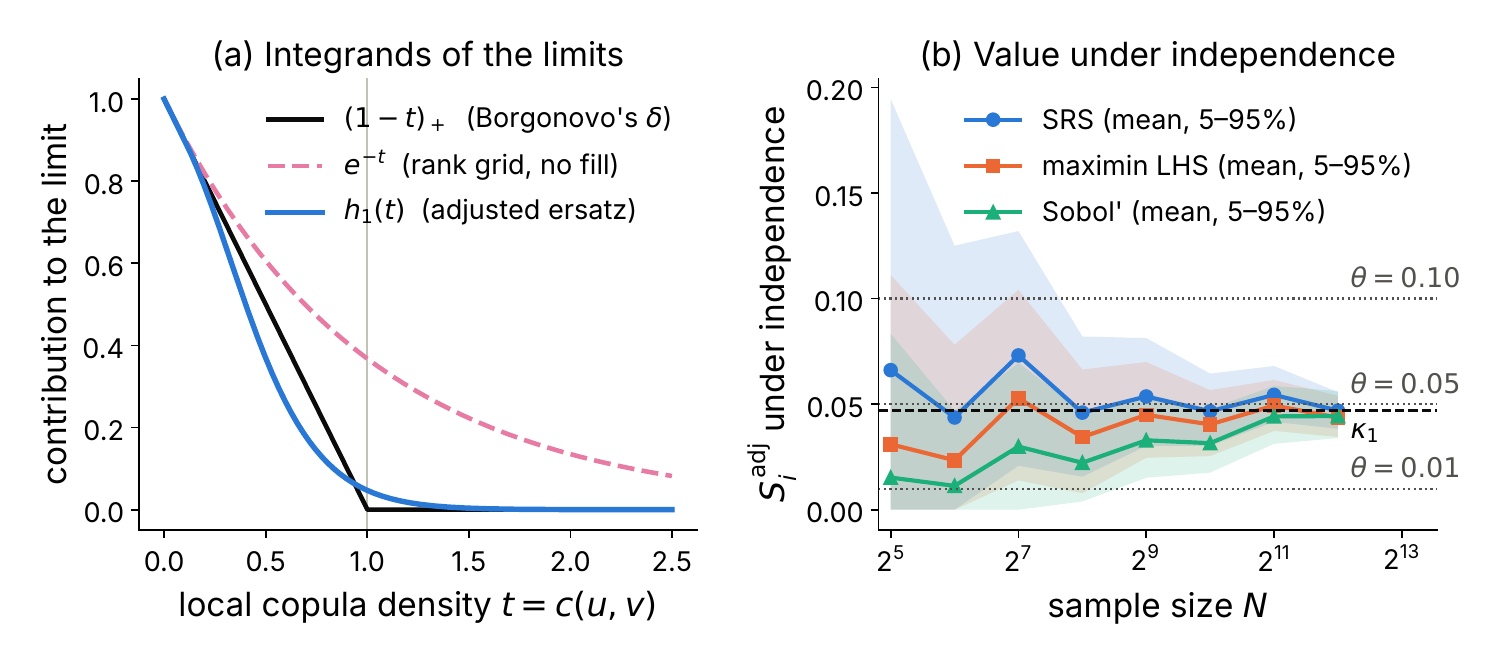}
    \caption{(a) Integrands of the three limit functionals as functions of the local copula density $t$: $(1-t)_+$ for $\delta_i$, $e^{-t}$ for the rank-transformed grid without fill ($\Phi_1$), and $h_1(t)$ for the adjusted ersatz ($\Psi_1$). (b) Adjusted ersatz of an input independent of the output (mean and 5\%--95\% band; output equal to the sum of five other inputs) under three sampling designs: scrambled Sobol' sequences and SRS (10{,}000 replicates), and maximin Latin hypercube samples (LHS; 300 to 2000 replicates); dotted lines mark the screening thresholds $\theta$ and the dashed line the limit $\kappa_1$. The oscillation with $N$ follows $N/\lceil\sqrt N\rceil^2$.}
    \label{fig:theory}
\end{figure}

Proposition~\ref{prop:hist} and Theorem~\ref{thm:limits} answer the question of what the ersatz estimates. The rank-transformed grid without fill is a histogram estimator of $\delta_i$, but with about one point per cell it is not consistent for $\delta_i$, which requires $N/s^2\to\infty$ (Supplementary Material A, Remark~5.3): it converges to $\Phi_1(C_i)$, which compresses the scale into $[0.37,1]$. The Moore fill acts as a local denoising step. In the limit it replaces the integrand $e^{-t}$ by $h_1(t)$, a smooth surrogate of the integrand $(1-t)_+$ of $\delta$ (Figure~\ref{fig:theory}a). The adjusted ersatz is thus a consistent estimator of a near-$\delta$ moment-independent measure, but not of $\delta_i$ itself, nor of $T_i$. The test that declares an input influential when the statistic exceeds $\kappa$ by a slowly vanishing margin is consistent against every alternative (Supplementary Material A, Corollary~4.5). The value under independence, $\kappa$, is positive. Figure~\ref{fig:theory}b shows the distribution of $\Sadj$ for an input independent of the output: at $N=512$ ($s=23$) its mean and standard deviation are $0.054$ and $0.015$ under SRS.

\begin{theorem}[grid resolution]\label{thm:grid}
Let $s=\lceil N^{\alpha_g}\rceil$ with $\alpha_g>1/4$. If $\alpha_g<1/2$, both statistics converge to the area of the region where $c_i=0$ (if its boundary has zero area), hence to zero for every copula with a positive density, whatever the strength of the dependence. If $\alpha_g>1/2$, both converge to one for every copula. Among these exponents, only $\alpha_g=1/2$ yields a non-degenerate measure of dependence. Coarser grids, with $\alpha_g\le1/4$, are not covered.
\end{theorem}

\begin{proposition}[relations with Sobol' indices]\label{prop:sobol}\leavevmode
\begin{enumerate}[label=(\alph*),leftmargin=*,itemsep=1pt]
\item If $T_i=0$, then $X_i$ and $Y$ are independent and all the statistics above converge to their values under independence.
\item The converse fails: for $Y=(X_1+X_2)\bmod1$, $X_1$ and $Y$ are independent although $T_1=1$.
\item If the output is additive in $X_i$ ($T_i=S_i$), then $Y-\mathbb E[Y\mid X_i]$ is independent of $X_i$, so a conditional spread that varies with $X_i$ signals interactions.
\end{enumerate}
\end{proposition}

\begin{theorem}[rankings]\label{thm:rank}
In single-index models $Y=g\big(\sum_ja_j\Phi_{\mathcal N}^{-1}(X_j)\big)$, with $\Phi_{\mathcal N}$ the standard normal distribution function and $g$ measurable with $0<\mathrm{Var}(Y)<\infty$, $T_i>T_j$ if and only if $|a_i|>|a_j|$, and likewise for $S_i$, while every $f$-sensitivity index, including $\delta_i$ and $\Phi_\lambda(C_i)-e^{-\lambda}$, is non-decreasing in $|a_i|$. Rankings by $T_i$ and by any $f$-sensitivity index therefore agree up to ties.
\end{theorem}

Outside this class the rankings by $T_i$ and $\delta_i$ can differ: for the Ishigami function $T_1>T_2$ but $\delta_2>\delta_1$ (Section~\ref{Benchmark}). $\Psi_\lambda$ is not an $f$-divergence, because $h_\lambda$ is not convex, and it equals $\kappa_\lambda$ rather than zero under independence. Although the adjusted ersatz is copula-based, its limit is therefore not one of the niveau~3 measures of \citet{plischke2019}, which vanish if and only if $X_i$ and $Y$ are independent (their Theorem~13, R\'enyi's postulate~D). Proposition~\ref{prop:limits}(b) is the corresponding property of $\Psi_\lambda$, relative to $\kappa_\lambda$. By Proposition~\ref{prop:limits}(c), $\Psi_\lambda$ ranks two inputs as $\delta$ does whenever their $\delta$ values differ by more than $0.19$. Numerically the sharp margin is $0.176$, and inputs whose values of $\delta$ differ by less than that can be ranked in either order (Supplementary Material A, Remark~7.3). Numerically, it is strictly increasing in the correlation of a Gaussian copula.

Theorem~\ref{thm:limits} is proved for SRS. For plain Latin hypercube samples the value under independence has the same limit (Supplementary Material A, Proposition~8.1), and the simulated values for maximin Latin hypercubes approach it (Figure~\ref{fig:theory}b). For scrambled Sobol' sequences the points are dependent and the value under independence depends on the model (Supplementary Material A, Section~8). When the output is driven by a single other input, the rank-transformed pairs inherit the regular two-dimensional structure of the Sobol' net, and the mean of the statistic of an independent input can fall to $0.0004$ at $N=512$, against $0.054$ under SRS. Conversely, projections of the sequence with poor uniformity produce regular patterns of empty cells. Such patterns are properties of the design and not evidence of interactions.

\subsection{Comparator estimators at equal sample cost}\label{comparators}

The ersatz statistics are compared with three estimators that use exactly the same input--output sample, at no additional model evaluations, and represent the approaches reviewed in the Introduction: an emulator-based estimator of $T_i$ (PCE), the Shapley effects derived from the same emulator, and a distribution-based index (PAWN). The metrics of the comparison are defined in Section~\ref{design}.

\paragraph*{Polynomial chaos expansion} The output is regressed on a full total-degree basis of orthonormal polynomials of the inputs: Legendre polynomials of the uniform inputs (inputs with other distributions are mapped to uniform ones through their distribution functions) and Hermite polynomials of the Gaussian inputs of the Oakley--O'Hagan function \citep{xiu2002}. The coefficients are obtained by ordinary least squares. The degree $p\in\{1,\dots,10\}$ is selected by minimising the corrected leave-one-out error \citep{blatman2010} among the degrees for which the number of terms $P$ satisfies $N\ge2P$ and $P\le1000$. Total-order indices follow from the squared coefficients $c_\alpha^2$ of the orthonormal basis, $T_i=\sum_{\alpha:\alpha_i>0}c_\alpha^2/\sum_{\alpha\neq0}c_\alpha^2$ \citep{sudret_global_2008}.

\paragraph*{Shapley effects} For independent inputs the Shapley effect of input $i$ is obtained from the same fit as $\mathrm{Sh}_i=\big(\sum_{\alpha:\alpha_i>0}c_\alpha^2/|\mathrm{supp}(\alpha)|\big)\big/\sum_{\alpha\ne0}c_\alpha^2$, where $\mathrm{supp}(\alpha)$ is the set of inputs on which the basis term depends. This allocates each variance component equally among the inputs involved \citep{owen2014}, and the effects sum to one. The permutation-based Monte Carlo estimator of \citet{song2016shapley} requires additional model runs and is not comparable at equal cost.

\paragraph*{PAWN} We use the given-data version of PAWN \citep{pianosiwagener2018}: the range of $x_i$ is divided into $n=10$ conditioning intervals of equal width, which are equiprobable for inputs on the probability scale, and the index is the maximum over the intervals of the Kolmogorov--Smirnov distance between the unconditional empirical distribution function of the output and the conditional one \citep{pianosiwagener2015, pianosiwagener2018}.

All estimators are implemented in Python with NumPy and SciPy \citep{virtanen2020} in a single module available in the repository (see the Data availability statement): the orthonormal bases, the least-squares fit and the Shapley allocation are our own code, PAWN uses \texttt{scipy.stats.ks\_2samp}, and the designs use \texttt{scipy.stats.qmc}.

\subsection{Experimental design}\label{design}

\paragraph*{Test cases and references} Table~\ref{tab:datasets} lists the test cases. Reference total-order indices are analytical for the Bratley--Fox (1988) product function (all equal to $0.278$), the Ishigami function and the Sobol' $G$ function. For the other functions they are Jansen estimates on scrambled Sobol' designs with base sample size $2^{17}$ (Bratley et al. 1992) or $2^{16}$ (Oakley--O'Hagan, Multiverse). Replicates with 14 other scrambling seeds agree with them to within $1\times10^{-3}$. The three continuous inputs of the Multiverse function have equal indices by symmetry. Reference values of $\delta_i$ are plug-in histogram estimates (Proposition~\ref{prop:hist}) with $64\times64$ equiprobable cells and $N=2^{24}$ points, about $4100$ points per cell, for which the residual bias is about $0.006$. Supplementary Material B lists all reference values.

\begin{table}[H]
\centering\small
\renewcommand{\arraystretch}{1.2}
\begin{tabular}{@{}>{\raggedright\arraybackslash}p{4.4cm}>{\raggedright\arraybackslash}p{8.9cm}@{}}\toprule
\textbf{Test case} & \textbf{Function and inputs}\\ \midrule
Bratley--Fox\newline{\footnotesize\citet{bratley1988algorithm}} & $y=\prod_{i=1}^{k}|4x_i-2|$\newline $k=8$; $x_i\sim\mathscr U[0,1]$\\ \addlinespace[3pt]
Bratley et al.\newline{\footnotesize\citet{bratley1992implementation}} & $y=\sum_{i=1}^{k}(-1)^i\prod_{j=1}^{i}x_j$\newline $k=8$; $x_i\sim\mathscr U[0,1]$\\ \addlinespace[3pt]
Ishigami\newline{\footnotesize\citet{ishigami1990importance}} & $y=\sin x_1+a\sin^2x_2+b\,x_3^4\sin x_1$\newline $k=3$; $x_i\sim\mathscr U[-\pi,\pi]$; $a=7$, $b=0.1$\\ \addlinespace[3pt]
Oakley--O'Hagan\newline{\footnotesize\citet{oakley2004probabilistic}} & $y=\bm a_1^\top\bm x+\bm a_2^\top\sin\bm x+\bm a_3^\top\cos\bm x+\bm x^\top\mathbf M\bm x$\newline $k=15$; $x_i\sim N(0,1)$; $\bm a_1$, $\bm a_2$, $\bm a_3$ and $\mathbf M$ as in the original article\\ \addlinespace[3pt]
Sobol' $G$\newline{\footnotesize\citet{sobol1998quasi}} & $y=\prod_{i=1}^{k}\dfrac{|4x_i-2|+a_i}{1+a_i}$\newline $k=8$; $x_i\sim\mathscr U[0,1]$; $\bm a=(0,1,4.5,9,99,99,99,99)$\\ \addlinespace[3pt]
Multiverse\newline{\footnotesize\citet{saltelli2025}} & $y=(x_1\circledast x_2\circledast x_3)^{1/3}$\newline $k=5$; $x_1,x_2,x_3$ uniform or truncated normal (switch $\zeta\in\{0,1\}$); the model form $\xi\in\{0,\dots,7\}$ sets the operations $\circledast$\\ \addlinespace[3pt]
Becker metafunction\newline{\footnotesize\citet{becker_metafunctions_2020}} & $\begin{aligned}[t]y={}&\textstyle\sum_{i}\alpha_if_i(x_i)+\sum_{(i,j)}\beta_{ij}f_i(x_i)f_j(x_j)\\[-1pt]&\textstyle+\sum_{(i,j,l)}\gamma_{ijl}f_i(x_i)f_j(x_j)f_l(x_l)\end{aligned}$\newline $k\in\{3,\dots,15\}$; see \S\ref{design}\\ \addlinespace[3pt]
HYMOD\newline{\footnotesize\citet{vrugt2003shuffled}} & Rainfall--runoff model\newline Output: Nash--Sutcliffe efficiency\newline $k=5$; uniform distributions (Table~\ref{tab:HYMOD})\\ \bottomrule
\end{tabular}
\caption{Test cases. Definitions follow the R package \texttt{sensobol} \citep{puy2022sensobol} and the original references. In the Multiverse function, $\xi$ selects one of eight combinations of sums and products of $x_1,x_2,x_3$, and $\zeta$ switches the three continuous inputs between $\mathscr U[0,1]$ and a normal distribution with mean $1/2$ and standard deviation $1/12$ truncated to $[0,1]$. In the Becker metafunction, $f_i$ is the univariate function drawn for input $i$, and the sums run over randomly selected pairs and triplets of inputs.}
\label{tab:datasets}
\end{table}

\paragraph*{Sample size, replicates and designs} All estimators are compared at $N=2^9=512$ model runs, a size at which the adjusted ersatz has stabilised and PCE is accurate (Section~\ref{convergence}). Each comparison is replicated 100 times, and we report medians over the replicates. Three designs are used: scrambled Sobol' sequences \citep{owen1995randomly}, as in \citet{puy2024discrepancy}; maximin Latin hypercube samples (LHS), the best of 20 random Latin hypercubes \citep{mckay2000} by minimum distance between points \citep{morris1995}; and SRS, whose points are independent and uniformly distributed (sometimes called pseudo-random sampling). The paper reports the Sobol' design, and Supplementary Material B the other two.

\paragraph*{Metrics} Agreement with a reference is measured by the Pearson correlation $\rho$ between the Savage scores of the estimates and of the reference values \citep{iman1987, puy2024discrepancy}, which emphasises agreement at the top of the ranking, and by the mean absolute error (MAE) across the inputs. We write $\rho_T$ and MAE$_T$ for the comparison with $T_i$, and $\rho_\delta$ and MAE$_\delta$ for the comparison with $\delta_i$. Only the PCE estimates of $T_i$ have one of the references as their target. For the other estimators, the ersatz statistics among them (\S\ref{theory}), the MAE measures the distance from a reference rather than an estimation error. Screening errors are measured at a threshold $\theta$. An input is truly influential if its reference $T_i$ exceeds $\theta$, and it is flagged if its estimate exceeds $\theta$. The type~I error $\alpha(\theta)$ is the share of non-influential inputs that are flagged, and the type~II error $\beta(\theta)$ the share of influential inputs that are not. $\alpha$ is undefined when no input has $T_i\le\theta$ and $\beta$ when none has $T_i>\theta$.

We use three thresholds, $\theta\in\{0.01,0.05,0.10\}$, which span conservative to liberal conventions for declaring an input non-influential. They are expressed on the scale of $T_i$, so they are natural for estimators of $T_i$. For the adjusted ersatz, whose value under independence is positive (Figure~\ref{fig:theory}b), they have a further meaning. Under SRS at $N=512$, an input independent of the output exceeds $\theta=0.01$, $0.05$ and $0.10$ with simulated probabilities $1.00$, $0.57$ and $0.004$, whatever the model (Supplementary Material A, Section~9): $\theta=0.01$ lies $2.8$ standard deviations below the mean of the statistic under independence ($0.054$), $\theta=0.05$ lies within $0.25$ standard deviations of it, and $\theta=0.10$ lies above its 99\% quantile ($0.093$). These probabilities apply to every input with $T_i=0$, because such an input is independent of the output (Proposition~\ref{prop:sobol}(a)). Inputs with a small positive $T_i$ behave in the same way only if their $\delta_i$ is also small, which a small $T_i$ does not guarantee (Supplementary Material A, Proposition~7.1). Under the Sobol' design the value under independence is lower and model-dependent (Section~\ref{theory}), and so are the type~I errors at $\theta=0.05$. A threshold on the statistic's own scale would be an upper quantile of its distribution under independence. Under SRS this distribution depends only on $N$, and Supplementary Material A tabulates it.

We report rankings as well as screening errors because screening in practice is often understood as the identification of the few most influential inputs, which is a statement about the top of the ranking. Convergence of the indices, of the rankings and of the screening results are also distinct properties \citep{sarrazin2016}.

\paragraph*{Convergence} For each function and for the Sobol' and SRS designs, 20 nested samples of size $N=2^4,\dots,2^{13}$ (the first $N$ points of one sequence) are evaluated with all estimators.

\paragraph*{Becker metafunction ensemble} We draw 512 configurations from a scrambled Sobol' design on $[0,1]^4$, mapped to the number of inputs $k\in\{3,\dots,15\}$, a base sample size $N_s\in\{10,\dots,100\}$, an input distribution family $\phi\in\{1,\dots,8\}$ (uniform, normal, four beta and a logit-normal distribution truncated to $[0,1]$, or one of them drawn per input) and a function seed $\epsilon\in\{1,\dots,200\}$ \citep{becker_metafunctions_2020, puy_comprehensive_2022}. The given-data estimators use $C=N_s(k+1)$ runs, between 40 and 1600, the cost of a Jansen design of base size $N_s$. Each configuration is evaluated with the three designs on the same function. The metafunction follows the implementation in \texttt{sensobol}: each input receives one of 13 univariate functions, coefficients are drawn from the mixture $0.7\,N(0,0.5)+0.3\,N(0,5)$, and $\lceil k/2\rceil$ pairwise and $\lceil k/5\rceil$ three-way interactions are included. Reference $T_i$ are Jansen estimates with base sample size $2^{13}$.

\paragraph*{Sensitivity analysis of the method} Five algorithmic parameters are varied jointly in a sensitivity analysis of the sensitivity analysis \citep{puy_sensitivity_2020}: the fill threshold $\theta_f\in[0.25,0.75]$; the grid exponent $\alpha_g\in[0.40,0.60]$, which sets the number of cells per axis as $s=\lceil N^{\alpha_g}\rceil$ (so $\alpha_g=1/2$ gives $s=\lceil\sqrt N\rceil$); the screening threshold $\theta\in[0.01,0.10]$; the number of PAWN conditioning intervals $n\in\{5,\dots,20\}$; and the sampling design (scrambled Sobol', maximin Latin hypercube or SRS). The five parameters are independent and uniformly distributed on these sets, as the Sobol' indices below require. The outputs are the means, over the first 128 configurations of the Becker ensemble, of $\rho_T$, MAE$_T$, $\alpha(\theta)$ and $\beta(\theta)$ for the adjusted ersatz and for PAWN. First-order and total-order indices of the five parameters are estimated from a scrambled Sobol' design of base size $2^{12}$ with the matrices $\mathbf A$, $\mathbf B$ and $\mathbf A_{\mathbf B}^{(i)}$ of \citet{saltelli_variance_2010} ($28{,}672$ evaluations), with the first-order estimator of \citet{saltelli_variance_2010} applied to the centred output, the total-order estimator of \citet{jansen1999}, and 95\% bootstrap intervals. Centring changes the expectation of the first-order estimator only by a term inversely proportional to the base sample size and removes from its variance a term proportional to the squared mean of the output.

\paragraph*{HYMOD} The HYMOD conceptual rainfall--runoff model \citep{vrugt2003shuffled} is applied to the Leaf River catchment (Mississippi) for the period from 28 July 1952 to 30 September 1962, with the Nash--Sutcliffe efficiency \citep[NSE;][]{nash1970river} as output. The sample consists of $50{,}000$ runs on an Owen-scrambled Sobol' sequence mapped to uniform distributions on the ranges of Table~\ref{tab:HYMOD}. Reference total-order indices are computed from the sample itself, so that the references and the estimates refer to the same runs: a PCE of degree seven (corrected leave-one-out error $0.1\%$ of the output variance) and a gradient-boosting regression \citep[test $R^2=0.998$;][]{pedregosa2011} were fitted to the $50{,}000$ runs, and the Jansen estimator was applied to each surrogate with base sample size $2^{15}$. The two sets of indices agree to within $0.002$. Because one surrogate is a PCE, agreement between PCE estimates and this reference is not an independent check. The $\delta_i$ reference is computed from the PCE surrogate with $2^{23}$ points and agrees to within $0.02$ with plug-in estimates computed directly from the sample on coarser grids.

\section{Results}\label{results}

\subsection{Convergence with sample size}\label{convergence}

\begin{figure}[H]
    \centering
    \includegraphics[width=\textwidth,height=0.88\textheight,keepaspectratio]{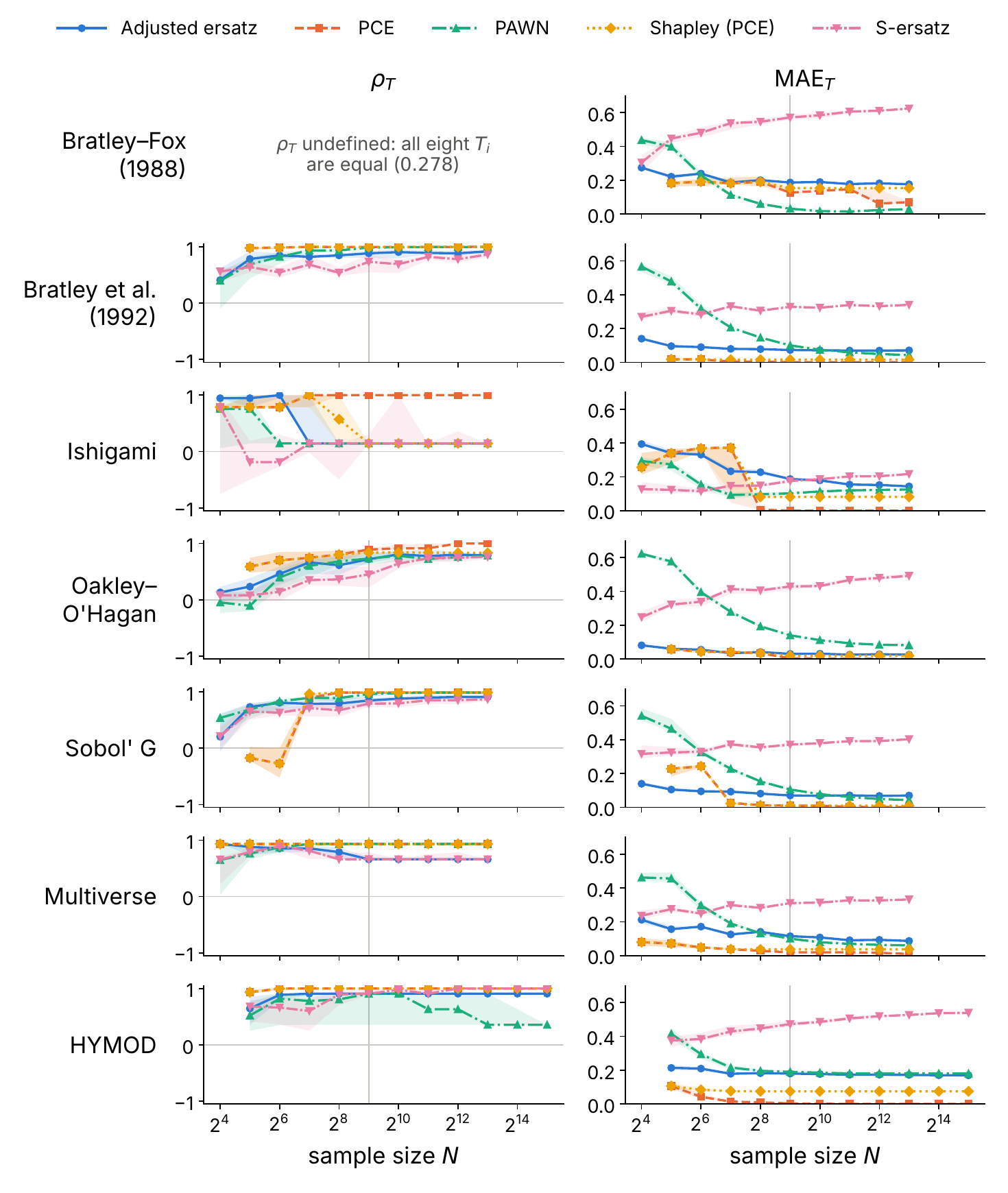}
    \caption{Convergence of $\rho_T$ (left) and MAE$_T$ (right) with the sample size $N$ for the five estimators (scrambled Sobol' design; median and inter-quartile band over 20 nested replicates; for HYMOD, blocks of the $50{,}000$-run sample). The vertical line marks $N=512$. The panels of each column share their axes, and $\rho_T$ is shown over its full range $[-1,1]$.}
    \label{fig:convergence}
\end{figure}

Figure~\ref{fig:convergence} shows the medians and inter-quartile ranges of $\rho_T$ and MAE$_T$ over 20 replicates as functions of $N$. The adjusted ersatz has essentially stabilised by $N=2^9=512$, the size used in the comparisons: its medians at $N=512$ are within $0.07$ ($\rho_T$) and $0.05$ (MAE$_T$) of their values at $N=8192$. Its behaviour beyond $N=512$ follows the theory. It converges to $\Psi_1(C_i)$, a quantity different from $T_i$, so its MAE$_T$ reaches a plateau and its ranking approaches the ranking of $\delta$. For the Multiverse function, where the rankings by $T_i$ and $\delta_i$ differ, its $\rho_T$ decreases from about $0.9$ at small $N$ to $0.66$ from $N=512$ on. The S-ersatz drifts upwards, as predicted for outputs with heavy tails (\S\ref{1st}). PCE becomes available once $N\ge2(k+1)$ and is the most accurate estimator of $T_i$ from $N=256$ on, except for the Bratley--Fox product function, whose absolute values are poorly represented by polynomials. The MAE$_T$ of PAWN decreases slowly with $N$, because its positive bias decreases with the number of points per conditioning interval. Supplementary Figure~S1 shows that the SRS design gives the same picture.

\subsection{Benchmark functions}\label{Benchmark}

\begin{figure}[H]
    \centering
    \includegraphics[width=\textwidth]{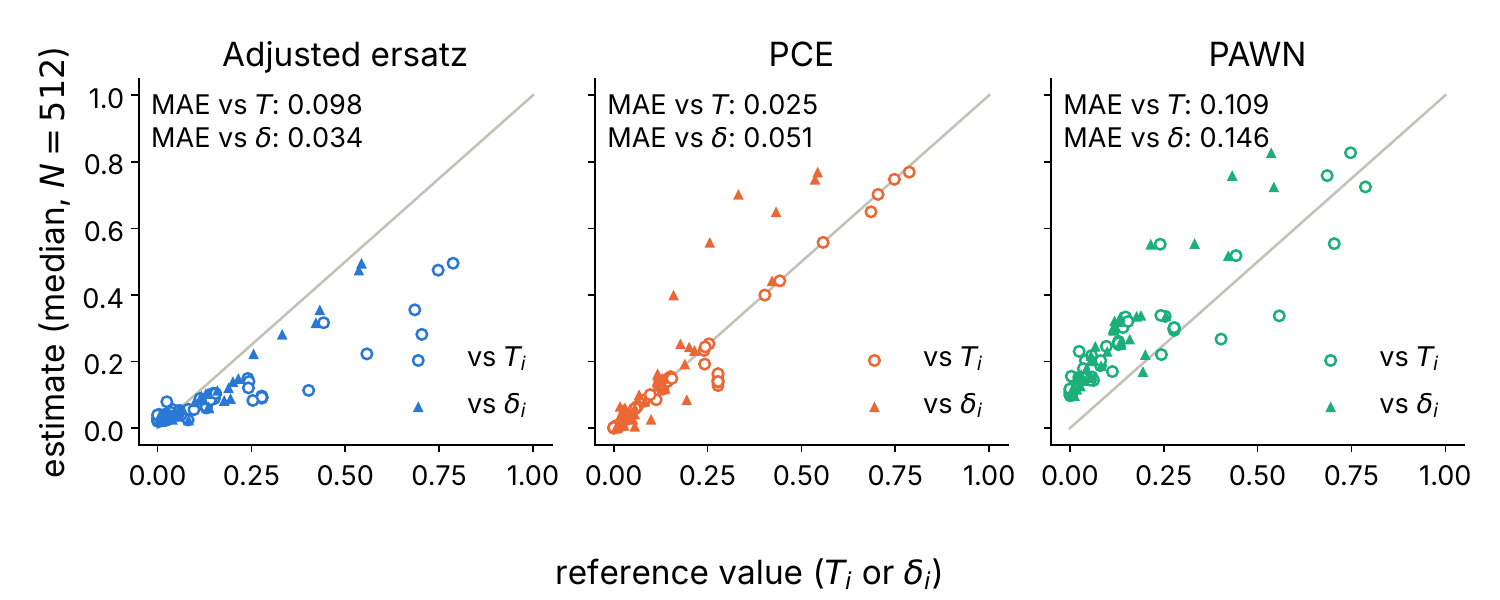}
    \caption{Median estimates at $N=512$ (scrambled Sobol' design) of every input of the six benchmark functions and HYMOD, plotted against the reference $T_i$ (open circles) and against the reference $\delta_i$ (filled triangles). Points on the diagonal agree with the reference. The adjusted ersatz tracks $\delta_i$, PCE tracks $T_i$, and PAWN overestimates both.}
    \label{fig:targets}
\end{figure}

Table~\ref{tab:agreement} reports the agreement of the five estimators with both references at $N=512$, and Table~\ref{tab:screening} the screening errors. Figure~\ref{fig:targets} plots the median estimates of all inputs against $T_i$ and against $\delta_i$.

\begin{table}[H]
\centering\footnotesize
\begin{tabular}{@{}llccccc@{}}\toprule
Test case & Metric & S-ersatz & Adj.\ ersatz & PCE & Shapley & PAWN \\ \midrule
Bratley--Fox (1988) & $\rho_T$ & -- & -- & -- & -- & -- \\
 & MAE$_T$ & 0.571 & 0.185 & 0.130 & 0.153 & \textbf{0.026} \\
 & $\rho_\delta$ & -- & -- & -- & -- & -- \\
 & MAE$_\delta$ & 0.732 & \textbf{0.025} & 0.048 & 0.031 & 0.185 \\
\addlinespace[2pt]
Bratley et al. (1992) & $\rho_T$ & 0.71 & 0.87 & \textbf{1.00} & \textbf{1.00} & 0.98 \\
 & MAE$_T$ & 0.329 & 0.074 & \textbf{0.001} & 0.016 & 0.100 \\
 & $\rho_\delta$ & 0.70 & 0.86 & \textbf{1.00} & \textbf{1.00} & 0.98 \\
 & MAE$_\delta$ & 0.327 & \textbf{0.035} & 0.042 & 0.036 & 0.131 \\
\addlinespace[2pt]
Ishigami & $\rho_T$ & 0.14 & 0.14 & \textbf{1.00} & 0.14 & 0.14 \\
 & MAE$_T$ & 0.181 & 0.186 & \textbf{0.001} & 0.081 & 0.107 \\
 & $\rho_\delta$ & \textbf{1.00} & \textbf{1.00} & 0.14 & \textbf{1.00} & \textbf{1.00} \\
 & MAE$_\delta$ & 0.302 & \textbf{0.065} & 0.122 & 0.093 & 0.067 \\
\addlinespace[2pt]
Oakley--O'Hagan & $\rho_T$ & 0.47 & 0.73 & \textbf{0.85} & 0.83 & 0.72 \\
 & MAE$_T$ & 0.430 & 0.031 & \textbf{0.004} & 0.019 & 0.142 \\
 & $\rho_\delta$ & 0.58 & 0.81 & 0.81 & \textbf{0.97} & 0.81 \\
 & MAE$_\delta$ & 0.447 & 0.021 & 0.021 & \textbf{0.009} & 0.159 \\
\addlinespace[2pt]
Sobol' $G$ & $\rho_T$ & 0.78 & 0.87 & \textbf{0.99} & \textbf{0.99} & 0.96 \\
 & MAE$_T$ & 0.378 & 0.072 & \textbf{0.010} & \textbf{0.010} & 0.106 \\
 & $\rho_\delta$ & 0.80 & 0.88 & \textbf{0.98} & \textbf{0.98} & 0.94 \\
 & MAE$_\delta$ & 0.372 & \textbf{0.031} & 0.037 & 0.037 & 0.120 \\
\addlinespace[2pt]
Multiverse & $\rho_T$ & 0.67 & 0.66 & \textbf{0.93} & \textbf{0.93} & \textbf{0.93} \\
 & MAE$_T$ & 0.310 & 0.111 & \textbf{0.019} & 0.037 & 0.103 \\
 & $\rho_\delta$ & \textbf{0.97} & \textbf{0.97} & 0.64 & 0.64 & 0.64 \\
 & MAE$_\delta$ & 0.315 & 0.080 & \textbf{0.075} & 0.091 & 0.142 \\
\addlinespace[2pt]
HYMOD & $\rho_T$ & 0.91 & 0.91 & \textbf{1.00} & \textbf{1.00} & 0.91 \\
 & MAE$_T$ & 0.473 & 0.180 & \textbf{0.004} & 0.075 & 0.191 \\
 & $\rho_\delta$ & 0.91 & \textbf{1.00} & 0.91 & 0.91 & \textbf{1.00} \\
 & MAE$_\delta$ & 0.575 & \textbf{0.040} & 0.150 & 0.096 & 0.181 \\
\addlinespace[2pt]
Becker ensemble & $\rho_T$ & 0.86 & 0.92 & \textbf{0.98} & \textbf{0.98} & 0.95 \\
 & MAE$_T$ & 0.393 & 0.064 & \textbf{0.011} & 0.012 & 0.146 \\
\bottomrule\end{tabular}
\caption{Agreement with the total-order reference ($\rho_T$, MAE$_T$) and with the $\delta$ reference ($\rho_\delta$, MAE$_\delta$) at $N=512$, scrambled Sobol' design: medians over 100 replicates (HYMOD: 97 disjoint blocks of the $50{,}000$-run sample; Becker ensemble: 512 configurations, sample sizes $40$ to $1600$). $\rho$ is the Pearson correlation of Savage scores (higher is better) and MAE the mean absolute error across inputs (lower is better). The best value in each row is in bold. $\rho$ is undefined for the Bratley--Fox function, whose eight inputs have equal $T_i$ and $\delta_i$. HYMOD references are computed from surrogates of the sample, one of them a PCE (\S\ref{design}).}
\label{tab:agreement}
\end{table}

\begin{table}[H]
\centering\footnotesize
\begin{tabular}{@{}llccccc@{}}\toprule
Test case & $\theta$ & S-ersatz & Adj.\ ersatz & PCE & Shapley & PAWN \\ \midrule
Bratley et al. (1992) & 0.01 & 1.00/0.00 & 1.00/0.00 & 0.00/0.00 & 0.00/0.00 & 1.00/0.00 \\
 & 0.05 & 1.00/0.00 & 0.00/0.33 & 0.00/0.00 & 0.00/0.00 & 1.00/0.00 \\
 & 0.10 & 1.00/0.00 & 0.00/0.50 & 0.00/0.00 & 0.00/0.00 & 0.83/0.00 \\
\addlinespace[2pt]
Oakley--O'Hagan & 0.05 & 1.00/0.00 & 0.50/0.27 & 0.00/0.00 & 0.00/0.36 & 1.00/0.00 \\
 & 0.10 & 1.00/0.00 & 0.00/0.60 & 0.10/0.00 & 0.00/0.00 & 1.00/0.00 \\
\addlinespace[2pt]
Sobol' $G$ & 0.01 & 1.00/0.00 & 1.00/0.00 & 0.00/0.25 & 0.00/0.25 & 1.00/0.00 \\
 & 0.05 & 1.00/0.00 & 0.17/0.00 & 0.00/0.00 & 0.00/0.00 & 1.00/0.00 \\
 & 0.10 & 1.00/0.00 & 0.00/0.00 & 0.00/0.00 & 0.00/0.00 & 0.83/0.00 \\
\addlinespace[2pt]
HYMOD & 0.01 & 1.00/0.00 & 1.00/0.00 & 0.00/0.00 & 0.00/0.00 & 1.00/0.00 \\
 & 0.05 & 1.00/0.00 & 0.50/0.00 & 0.00/0.00 & 0.00/0.00 & 1.00/0.00 \\
 & 0.10 & 1.00/0.00 & 0.00/0.00 & 0.00/0.00 & 0.00/0.00 & 1.00/0.00 \\
\addlinespace[2pt]
Becker ensemble & 0.01 & 1.00/0.00 & 0.91/0.01 & 0.07/0.07 & 0.04/0.09 & 1.00/0.00 \\
 & 0.05 & 1.00/0.00 & 0.23/0.07 & 0.02/0.08 & 0.02/0.10 & 1.00/0.00 \\
 & 0.10 & 1.00/0.00 & 0.04/0.15 & 0.02/0.08 & 0.02/0.09 & 0.86/0.00 \\
\bottomrule\end{tabular}
\caption{Screening errors $\alpha(\theta)/\beta(\theta)$ at $N=512$, scrambled Sobol' design (medians over replicates; Becker ensemble: means over configurations). $\alpha(\theta)$ is the share of the inputs with $T_i\le\theta$ whose estimate exceeds $\theta$ (type~I error), and $\beta(\theta)$ the share of the inputs with $T_i>\theta$ whose estimate does not (type~II error). Rows where $\alpha$ is undefined (no input with $T_i\le\theta$) are omitted; the full table, including the $\beta$ values of the Bratley--Fox, Ishigami and Multiverse functions, is in Supplementary Material B.}
\label{tab:screening}
\end{table}

Three results stand out. First, PCE and the adjusted ersatz follow different references, as their targets predict (Figure~\ref{fig:targets}). PCE, with its degree selected by cross-validation, estimates $T_i$ accurately on every function: $\rho_T\ge0.85$ and MAE$_T\le0.02$, except for the Bratley--Fox product function. The adjusted ersatz, whose target lies near $\delta_i$, has the smallest MAE$_\delta$ on five of the seven test cases, between $0.025$ and $0.065$, and is close to the best on the other two (Oakley--O'Hagan: $0.021$ against $0.009$ for the Shapley effects; Multiverse: $0.080$ against $0.075$ for PCE). MAE$_\delta$ ranges from $0.021$ to $0.150$ for PCE and from $0.067$ to $0.185$ for PAWN. Across all inputs of the benchmark functions and HYMOD, the median values of the adjusted ersatz differ from $\delta_i$ by $0.034$ on average and from $T_i$ by $0.098$, as predicted by Proposition~\ref{prop:limits}(c).

Second, where the rankings by $T_i$ and by $\delta_i$ differ, the ersatz statistics follow $\delta$. For the Ishigami function $T_1=0.558>T_2=0.442$, while $\delta_2=0.42>\delta_1=0.26$: the conditional distribution of the output changes more with $x_2$, through the large main effect $7\sin^2x_2$, than with $x_1$, whose total effect partly comes from its interaction with $x_3$. All distribution-based estimators, and the Shapley effects, which split the interaction equally between $x_1$ and $x_3$, rank $x_2$ first and obtain $\rho_T=0.14$ and $\rho_\delta=1$. For the Multiverse function the distributional switch $\zeta$ has a smaller $T_i$ ($0.113$) but a larger $\delta_i$ ($0.194$) than the continuous inputs ($T_i=0.130$, $\delta_i=0.136$), and the ersatz statistics again follow $\delta$ ($\rho_\delta=0.97$). These are not estimation errors but differences between the variance-based and the moment-independent notions of importance (Theorem~\ref{thm:rank} and the discussion that follows it).

Third, the two modifications improve the S-ersatz substantially. The adjusted ersatz raises $\rho_T$ for the Bratley et al., Oakley--O'Hagan and Sobol' $G$ functions and in the Becker ensemble, leaves it unchanged for the Ishigami function and HYMOD, and lowers it marginally for the Multiverse function ($0.66$ against $0.67$). It divides MAE$_T$ by a factor of $2.6$ to $14$ on every test case except the Ishigami function, where both are about $0.18$, and MAE$_\delta$ by a factor of $4$ to $30$. The S-ersatz is far from both references (MAE$_T$ between $0.18$ and $0.57$, MAE$_\delta$ between $0.30$ and $0.73$) because of its dependence on the output distribution (\S\ref{1st}).

The type~I errors of the adjusted ersatz in Table~\ref{tab:screening} reflect its value under independence (\S\ref{design}). It flags every non-influential input of the benchmark functions at $\theta=0.01$ (91\% on average in the Becker ensemble), a threshold that even inputs independent of the output exceed. At $\theta=0.05$ it flags half of them for Oakley--O'Hagan and HYMOD, whose non-influential inputs have small but positive $\delta_i$, and fewer ($0.17$ and $0.00$) for the Sobol' $G$ and Bratley et al. functions, whose output is dominated by one input, so that the Sobol' design lowers the value under independence. At $\theta=0.10$ its median type~I error is zero on every benchmark function, and its mean is $0.04$ in the Becker ensemble. Its type~II errors arise for inputs with $T_i>\theta$ whose limit $\Psi_\lambda(C_i)$ lies below $\theta$ or within sampling error of it. At $\theta=0.05$ this happens for the Oakley--O'Hagan inputs $x_1$, $x_2$ and $x_4$, whose total-order indices ($0.055$--$0.063$) come almost entirely from interactions (first-order indices of at most $0.003$): their $\delta_i$ are $0.017$ to $0.030$, and their limits are close to the value under independence. At $\theta=0.10$ it happens for $x_{11}$, $x_{13}$ and $x_{14}$ of the same function ($T_i\approx0.14$, $\delta_i\approx0.12$), whose limits at $N=512$, about $0.108$, exceed the threshold by less than one standard deviation of the statistic (about $0.02$). The Sobol' design adds to these errors. When one input dominates the output, it lowers the statistic of every input, not only of the non-influential ones: for $x_2$ and $x_3$ of the Bratley et al.\ (1992) function the means are $0.08$ and $0.03$, against $0.14$ and $0.08$ under SRS, and these inputs account for its type~II errors at $\theta=0.10$ and $0.05$.

Under SRS, which the theory covers, the screening errors of the adjusted ersatz agree with it quantitatively (Supplementary Material B, Section~B.7). The 345 inputs of the Becker ensemble with $T_i=0$ are flagged at the rates of the distribution under independence at their sample sizes ($0.617$ against $0.608$ at $\theta=0.05$, $0.026$ against $0.025$ at $\theta=0.10$). In the convergence study the type~I error at $\theta=0.05$ oscillates with $N/s^2$ as that distribution does, and it does not vanish as $N$ grows, because the limits of inputs with a small positive $\delta_i$ exceed $0.05$. The limits of all 47 benchmark inputs lie within the bounds of Proposition~\ref{prop:limits}(c), and the 19 pairs of benchmark inputs whose $\delta_i$ differ by more than $0.19$ are ranked as by $\delta$ in every replicate of the three designs.

PCE makes few screening errors. PAWN flags most non-influential inputs at every threshold (type~I errors of at least $0.83$), because its value for an independent input, the maximum of ten Kolmogorov--Smirnov distances computed on about 51 points each, is about $0.12$ (Sobol') to $0.18$ (SRS) at $N=512$ when the output is the sum of five other inputs.

Supplementary Material B reports the results for maximin Latin hypercube and SRS designs. Pooled over the benchmark functions, the median $\rho_T$ of the adjusted ersatz is $0.76$ (Sobol'), $0.79$ (Latin hypercube) and $0.79$ (SRS), and its MAE$_\delta$ is $0.034$, $0.032$ and $0.035$: the conclusions do not depend on the design. The design matters more for PAWN, whose median $\rho_T$ drops from $0.93$ (Sobol') to $0.87$ (SRS).

\subsection{Becker metafunction ensemble}\label{BeckerSA}

\begin{figure}[H]
    \centering
    \includegraphics[width=\textwidth]{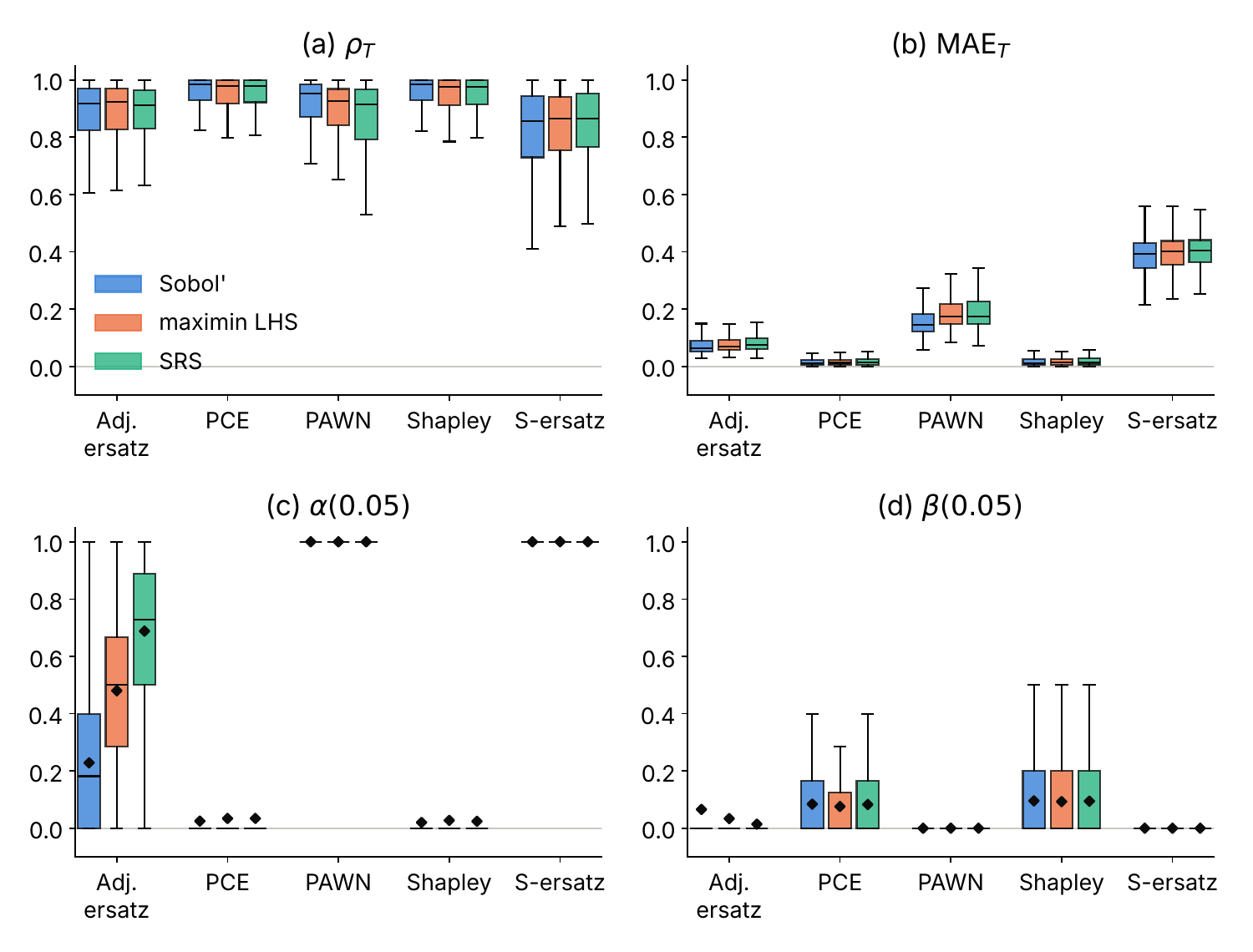}
    \caption{Distributions of (a) $\rho_T$, (b) MAE$_T$, (c) $\alpha(0.05)$ and (d) $\beta(0.05)$ over the 512 configurations of the Becker ensemble for the five estimators and three designs. All panels have the same vertical scale. Boxes: inter-quartile ranges and medians; whiskers: most extreme values within 1.5 inter-quartile ranges of the box (values beyond them are not shown); diamonds: means of $\alpha$ and $\beta$.}
    \label{fig:becker}
\end{figure}

The Becker ensemble removes the selection of test functions and includes discontinuous and piecewise univariate functions, small samples (40 to 1600 runs) and up to fifteen inputs, and 42\% of its inputs have $T_i\le0.01$. Table~\ref{tab:becker_function} and Figure~\ref{fig:becker} summarise the results. PCE and Shapley effects agree best with $T_i$ (median $\rho_T=0.98$, MAE$_T\approx0.012$), even though the selected degree is one or two in three quarters of the configurations. The adjusted ersatz ranks the inputs better than the S-ersatz (median $\rho_T$ of $0.91$--$0.92$ against $0.86$--$0.87$) and has a six times smaller MAE$_T$ ($0.064$--$0.075$ against $0.39$--$0.40$). PAWN ranks as well as or better than the adjusted ersatz (median $\rho_T$ of $0.91$--$0.95$) but with more than twice its MAE$_T$. The adjusted ersatz ranks the inputs better than PCE in 18\% to 20\% of the configurations, and PCE has $\rho_T<0.8$ in 10\% to 13\% of them, against about 20\% for the adjusted ersatz. The type~I error of the adjusted ersatz at $\theta=0.05$ depends on the design (mean $0.23$ for Sobol', $0.48$ for Latin hypercube and $0.69$ for SRS), as expected from the design dependence of its value under independence (Figure~\ref{fig:theory}b).

\begin{table}[H]
\centering\footnotesize
\begin{tabular}{@{}llccccc@{}}\toprule
Design & Metric & S-ersatz & Adj.\ ersatz & PCE & Shapley & PAWN \\ \midrule
Sobol' & $\rho_T$ & 0.86 & 0.92 & \textbf{0.98} & \textbf{0.98} & 0.95 \\
 & MAE$_T$ & 0.393 & 0.064 & \textbf{0.011} & 0.012 & 0.146 \\
 & $\bar\alpha/\bar\beta$ (0.05) & 1.00/0.00 & 0.23/0.07 & 0.02/0.08 & 0.02/0.10 & 1.00/0.00 \\
\addlinespace[2pt]
Maximin LHS & $\rho_T$ & 0.87 & 0.92 & \textbf{0.98} & \textbf{0.98} & 0.93 \\
 & MAE$_T$ & 0.401 & 0.070 & \textbf{0.012} & 0.013 & 0.174 \\
 & $\bar\alpha/\bar\beta$ (0.05) & 1.00/0.00 & 0.48/0.03 & 0.03/0.08 & 0.03/0.09 & 1.00/0.00 \\
\addlinespace[2pt]
SRS & $\rho_T$ & 0.87 & 0.91 & \textbf{0.98} & \textbf{0.98} & 0.91 \\
 & MAE$_T$ & 0.404 & 0.075 & \textbf{0.013} & 0.013 & 0.175 \\
 & $\bar\alpha/\bar\beta$ (0.05) & 1.00/0.00 & 0.69/0.01 & 0.03/0.08 & 0.02/0.09 & 1.00/0.00 \\
\bottomrule\end{tabular}
\caption{Becker metafunction ensemble (512 configurations, each evaluated with the three designs): medians of $\rho_T$ and MAE$_T$, and means of the screening errors at $\theta=0.05$.}
\label{tab:becker_function}
\end{table}

\subsection{Sensitivity analysis of the method}\label{SoS}

\begin{figure}[H]
    \centering
    \includegraphics[width=\textwidth]{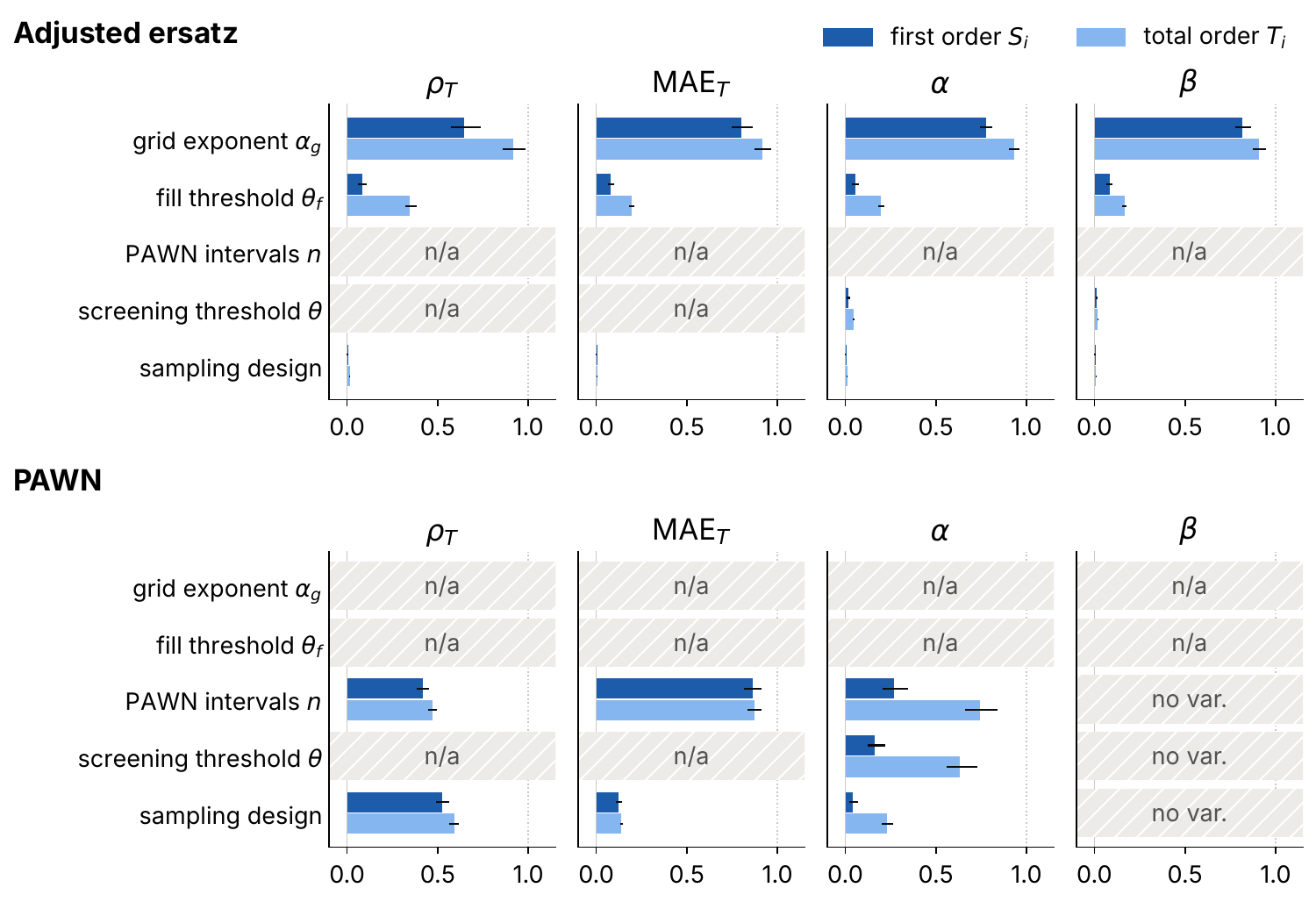}
    \caption{First-order ($S_i$, dark) and total-order ($T_i$, light) indices, with 95\% bootstrap intervals, of the five algorithmic parameters on the performance metrics of the adjusted ersatz (top) and of PAWN (bottom), averaged over 128 configurations of the Becker ensemble. The difference $T_i-S_i$ is the contribution of interactions. Hatched: the parameter does not enter the metric (n/a), or the metric has negligible variance (no var.).}
    \label{fig:sos}
\end{figure}

Figure~\ref{fig:sos} reports the first-order and total-order indices of the five algorithmic parameters, with structurally inactive combinations marked as not applicable: the fill threshold and the grid exponent do not enter PAWN, the number of PAWN intervals does not enter the adjusted ersatz, and the screening threshold enters only the screening errors. The type~II error of PAWN is zero in 96\% of the evaluations and its maximum is $0.017$, so its variance is too small for the indices to be meaningful. Following \citet{sarrazin2016}, we checked convergence on the nested designs of base size $2^6$ to $2^{12}$ (Supplementary Material~B, Figure~S2). By their criteria the ranking and the screening of the parameters have converged at $2^{12}$ for every metric, but not the values of the larger indices, whose 95\% intervals are up to $0.18$ wide against a threshold of $0.05$. These values change by at most $0.023$ between base sizes $2^{11}$ and $2^{12}$.

The grid exponent dominates every metric of the adjusted ersatz ($T_i\approx0.9$), followed by the fill threshold ($T_i$ from $0.17$ to $0.35$). This is Theorem~\ref{thm:grid} at work: at fixed $N$ the ratio $N/s^2$ ranges from about $N^{-0.2}$ to $N^{0.2}$ over $\alpha_g\in[0.40,0.60]$, and the value of the statistic under independence moves from about $0.75$ to $2\cdot10^{-9}$, which shifts all its values and its screening errors. On the finest grids ($0.0422<N/s^2<0.4815$, as for $\alpha_g>0.548$ when $N=1600$), dependence can even lower the statistic below its value under independence (Supplementary Material A, Proposition~5.4). The first-order indices show how the two parameters act. Most of the effect of the grid exponent is direct ($S_i$ from $0.65$ to $0.82$), whereas $50\%$ to $75\%$ of the total-order index of the fill threshold comes from its interaction with the grid exponent. For $\rho_T$, for instance, its first-order index is $0.09$ against a total-order index of $0.35$, a difference of $0.26$ (95\% bootstrap interval $[0.23,0.30]$). The interaction is expected. The two parameters are sampled independently, but the statistic responds to them jointly: the fill changes the limit through $h_\lambda$ (Theorem~\ref{thm:limits}), and the effect of $\theta_f$ on $h_\lambda$ varies with $\lambda=N/s^2$, which the grid exponent sets. Over $\theta_f\in[0.25,0.75]$, the value of the statistic under independence ranges from $0.14$ to $0.66$ at $\lambda=0.4$ and from $0.002$ to $0.23$ at $\lambda=1$, but only from $0$ to $0.04$ at $\lambda=1.6$, because on coarse grids few cells are empty.

The sampling design has a small effect on the adjusted ersatz ($T_i\le0.02$) but a large one on the ranking accuracy of PAWN ($T_i=0.59$), whose metrics are otherwise driven by the number of conditioning intervals and, for the screening errors, by the threshold. The ranking accuracy and the mean absolute error of PAWN are close to additive in these parameters, with first-order indices summing to $0.94$ and $0.99$. Its type~I error is not. It exceeds $0.9$ in 94\% of the evaluations and falls below this value only when a high threshold is combined with few conditioning intervals, most often under the Sobol' design, so the three parameters act mainly through their interactions (sum of first-order indices $0.47$). 

\subsection{Real-world application: the HYMOD hydrological model}\label{HYMOD}

Table~\ref{tab:HYMOD} lists the five HYMOD parameters, which are uniformly distributed on the ranges given. The NSE output is bounded above by one, strongly left-skewed (skewness $-3.7$) and negative in 27\% of the runs.

\begin{table}[H]
\centering\footnotesize
\begin{tabular}{@{}llp{6.4cm}l@{}} \toprule
\textbf{Parameter} & \textbf{Distribution} & \textbf{Description} & $T_i$ / $\delta_i$\\ \midrule
Cmax  & $\mathscr U[1,500]$ & Maximum storage capacity of the catchment [mm] & 0.241 / 0.215\\
bexp  & $\mathscr U[0.1,2]$ & Spatial variability of the soil moisture capacity [--] & 0.025 / 0.100\\
alpha & $\mathscr U[0.1,0.99]$ & Factor distributing the flow between the quick- and slow-flow reservoirs [--] & 0.403 / 0.159\\
Rs    & $\mathscr U[0,0.1]$ & Recession (residence-time) parameter of the slow-flow reservoir [--] & 0.004 / 0.056\\
Rq    & $\mathscr U[0.1,0.99]$ & Recession (residence-time) parameter of the quick-flow reservoirs [--] & 0.704 / 0.332\\ \bottomrule
\end{tabular}
\caption{HYMOD parameters, their uniform distributions \citep{vrugt2003shuffled}, and the reference indices computed from the $50{,}000$-run sample (\S\ref{design}).}
\label{tab:HYMOD}
\end{table}

The quick-flow parameter Rq dominates the NSE ($T_i=0.70$), followed by alpha and Cmax. The slow-flow parameter Rs has almost no influence on the output variance ($T_i=0.004$). This is consistent with the sensitivity of the NSE to peak flows, which are controlled by the quick-flow reservoirs. The rankings by $T_i$ and by $\delta_i$ differ: Cmax shifts the bulk of the NSE distribution more than alpha does ($\delta_i=0.215$ against $0.159$), while alpha accounts for more of its variance, which is dominated by the long lower tail.

\begin{table}[H]
\centering\footnotesize
\begin{tabular}{@{}lccccc@{}}\toprule
Metric & S-ersatz & Adj.\ ersatz & PCE & Shapley & PAWN \\ \midrule
$\rho_T$ & \textbf{1.00} & 0.91 & \textbf{1.00} & \textbf{1.00} & 0.36 \\
MAE$_T$ & 0.550 & 0.171 & \textbf{0.000} & 0.075 & 0.180 \\
$\rho_\delta$ & 0.91 & \textbf{1.00} & 0.91 & 0.91 & 0.63 \\
MAE$_\delta$ & 0.653 & \textbf{0.022} & 0.153 & 0.095 & 0.139 \\
$\alpha/\beta$ (0.01) & 1.00/0.00 & 1.00/0.00 & 0.00/0.00 & 0.00/0.00 & 1.00/0.00 \\
$\alpha/\beta$ (0.05) & 1.00/0.00 & 0.50/0.00 & 0.00/0.00 & 0.00/0.00 & 1.00/0.00 \\
$\alpha/\beta$ (0.10) & 1.00/0.00 & 0.00/0.00 & 0.00/0.00 & 0.00/0.00 & 0.50/0.00 \\
\bottomrule\end{tabular}
\caption{HYMOD: estimators computed on the full sample of $50{,}000$ runs. The reference $T_i$ were obtained from surrogates of this sample, one of them a PCE, so the perfect agreement of PCE with $T_i$ is partly by construction (\S\ref{design}).}
\label{tab:hymod_summary}
\end{table}

On the full sample (Table~\ref{tab:hymod_summary}) the adjusted ersatz ranks the parameters exactly as $\delta$ does ($\rho_\delta=1$), and its values differ from $\delta_i$ by $0.022$ on average (MAE$_\delta$). Against $T_i$ it swaps alpha and Cmax ($\rho_T=0.91$). The S-ersatz ranks the parameters as $T_i$ does here, but its values ($0.81$--$0.86$ for every parameter) are uninformative about magnitude, as expected for a strongly skewed output (\S\ref{1st}). PCE reproduces the surrogate-based reference by construction, and the gradient-boosting surrogate confirms these indices independently. PAWN gives similar values to Cmax and Rq ($0.53$ against $0.53$) and its ranking against $T_i$ deteriorates as $N$ grows (Figure~\ref{fig:convergence}), because with many points per interval it orders the parameters by the largest local distribution shift, which is not the ordering of $T_i$. At $N=512$ (Table~\ref{tab:agreement}) the adjusted ersatz again ranks as $\delta$ does in the median replicate, with MAE$_\delta=0.040$, and it reaches this ranking from $N=256$ on.

\section{Discussion and conclusions}\label{disc}

This paper set out to provide a screening statistic for the given-data setting whose properties are established before it is applied. The adjusted ersatz extends the S-ersatz of \citet{puy2024discrepancy} with a rank transformation of the output and a Moore-neighbourhood fill, and the theory of Section~\ref{theory} states what it estimates. The two modifications have distinct roles. The rank transformation removes the dependence of the statistic on the output distribution, which makes the S-ersatz drift towards one for skewed or heavy-tailed outputs. The fill changes what the rank-transformed grid estimates. Without the fill, the grid is a histogram estimator of Borgonovo's $\delta$, but at the resolution of the ersatz, about one point per cell, this estimator does not converge to $\delta$. With the fill, it is a consistent estimator of a functional that stays between $\delta-0.141$ and $\delta+0.048$ for every dependence structure. \citet{plischke2019} estimate niveau~3 measures, $\delta$ among them, by smoothing the empirical copula with Bernstein polynomials. For $\delta$ their estimator is consistent when the copula has continuous third derivatives and the polynomial degree grows suitably with the sample size (their Theorem~16). The fill plays the role of this smoothing at a resolution of about one point per cell. It requires less regularity (Section~\ref{theory}) and costs little more than sorting the output, but its target is a near-$\delta$ measure rather than $\delta$ itself. When $\delta$ itself is required, an estimator such as theirs is the appropriate tool.

Three consequences matter for practice. First, the adjusted ersatz is a moment-independent measure, and it should be compared with $T_i$ with this in mind. When the variance-based and the moment-independent notions of importance agree, as they do in single-index models (Theorem~\ref{thm:rank}) and approximately on most of our test cases, the adjusted ersatz ranks the inputs as $T_i$ does. When they disagree, as for the Ishigami and Multiverse functions and for HYMOD, it followed $\delta$ in our experiments, as the theory guarantees when the $\delta$ values of the inputs are well separated (Section~\ref{theory}). Both notions are legitimate: $T_i$ measures the share of output variance attributable to an input and has a direct interpretation for factor fixing \citep{sobol2001}, while $\delta_i$ measures how much the input shifts the whole output distribution, which matters when the variance is not a good summary of uncertainty \citep{borgonovo2011, baroni2020}. For screening, the question whether an input can be fixed, the two coincide on one side: an input with $T_i=0$ is independent of the output and has $\delta_i=0$, so it is non-influential under both. The only inputs that a bivariate statistic can miss are those whose interactions leave no trace on the conditional distribution of the output (Proposition~\ref{prop:sobol}).

Second, the value of the adjusted ersatz under independence, about $0.05$ at the default resolution, sets a floor below which no threshold can separate influential from non-influential inputs. The type~I errors at $\theta=0.01$, $0.05$ and $0.10$ observed throughout Section~\ref{results} are the direct consequence of this floor and not an idiosyncrasy of particular functions. Thresholds for the adjusted ersatz should therefore be set on its own scale, for instance as an upper quantile of its distribution under independence, which under simple random sampling depends only on $N$ and is tabulated in Supplementary Material A.

Third, when a polynomial surrogate is adequate, PCE estimates $T_i$ far more accurately than any distribution-based statistic at the same cost. With its degree selected by cross-validation, PCE was accurate on all but one of our analytical test cases and on the HYMOD sample, and its leave-one-out error provides the diagnostic that tells the analyst whether the surrogate can be trusted. The adjusted ersatz does not compete with PCE on that ground. Its role is different: it requires no functional-form assumption, no basis or degree selection and no fitting, it is computable for any sample size and any input type, it costs little more than sorting the output, and its behaviour is characterised by a theorem rather than by a benchmark. Among the distribution-based given-data statistics we compared, its values were the closest to $\delta$. They were also closer to $T_i$ than the S-ersatz on every test case except the Ishigami function, where both MAE$_T$ are about $0.18$, and closer than PAWN on four of the seven test cases and in the Becker ensemble (median MAE$_T$ of $0.064$ against $0.146$).

The joint sensitivity analysis of the method's parameters shows that the grid resolution matters most, as Theorem~\ref{thm:grid} predicts: only $\alpha_g=1/2$ yields a non-degenerate limit, and it should be kept as the default. The fill threshold is second in importance and acts in large part through its interaction with the grid resolution, and the sampling design has little effect on the adjusted ersatz.

These results suggest the following use. When the model is smooth enough for a polynomial surrogate, as indicated by a small leave-one-out error, PCE is the estimator of choice for $T_i$. The adjusted ersatz is a fast, assumption-free first screen of a given-data sample, with thresholds calibrated on its own scale, and a check on surrogate-based indices. When the distribution of the output rather than its variance is of interest, its values also approximate $\delta$ within known bounds. Its result can be explained to a non-specialist with a single scatterplot (Figure~\ref{fig:fill}), and it now rests on an explicit statement of what it estimates.

Five directions for further work follow from the limitations identified here. The theory is proved for simple random sampling, and a limit theory for scrambled nets, whose behaviour we documented only numerically, would clarify when quasi-random designs help. Thresholds calibrated on the null distribution deserve a systematic evaluation. The grid size could be tuned: for $N/s^2\le0.79$ the limit of the adjusted ersatz never falls below $\delta$, at the price of a higher value under independence (about $0.12$ at $N/s^2=0.79$; Supplementary Material A, Proposition~5.4). The interaction index of \citet{baroni2020} suggests a further use of the statistic. Computed on the pairs $(x_i^{(j)},\,y_j-\hat m_i(x_i^{(j)}))$, where $\hat m_i$ estimates the conditional mean of the output, the adjusted ersatz would measure the dependence left once the first-order effect is removed. Since the centred output is independent of $x_i$ when the output is additive in $x_i$ (Proposition~\ref{prop:sobol}(c)), its excess over the value under independence would indicate interactions, without kernel density estimation. Finally, interactions that leave no trace on bivariate projections could be detected by extending the occupancy idea to trivariate projections $(x_i,x_l,Y)$, at a cost of $O(k^2)$ rather than $O(k)$ grid evaluations.

\paragraph*{Acknowledgements} We thank the three reviewers for their comments. The observation of one of them that the rank-transformed statistic is a histogram estimator of Borgonovo's $\delta$ led to the theory of Section~\ref{theory}.

\paragraph*{Disclosure statement} The authors declared no potential conflicts of interest concerning the research, authorship, and/or publication of this article.

\paragraph*{Funding statement} The authors received no specific funding for this work.

\paragraph*{Data availability statement}\sloppy The original R implementation of \citet{puy2024discrepancy} and the \href{https://cran.r-project.org/web/packages/sensobol/index.html}{\texttt{sensobol}} package are available at \href{https://github.com/arnaldpuy/sensobol}{https://github.com/arnaldpuy/sensobol}. The Python code of this paper (estimators, test functions, experiments, figures), the HYMOD sample and the supplementary material are available at \href{https://github.com/Confareneoclassico/Discrepancy}{https://github.com/Confareneoclassico/Discrepancy}. Release 2.0.0, used for this paper, is archived on Zenodo (\href{https://doi.org/10.5281/zenodo.24173619}{https://doi.org/10.5281/zenodo.24173619}).

\paragraph*{Declaration of generative AI in the writing process} During the preparation of this work, the authors used a large language model assistant to support code checking and editing of the manuscript text. The authors reviewed and edited all content and take full responsibility for the content of the published article.

\paragraph*{Authors' contributions} S.L.P., A.L., and A.S. conceptualized the project and wrote the paper; S.L.P. and A.L. wrote the code and conducted the statistical analysis; R.S. supported the hydrological model analysis; A.P. and P.T.R. provided critical feedback; A.S. supervised the project. All the authors read and approved the final version of the paper.

\bibliographystyle{elsarticle-harv}
\bibliography{biblio}

\end{document}